\newif\ifShowKeys
\ShowKeysfalse

\documentclass[11pt,a4paper]{article} 
\pdfoutput=1
\usepackage[no-natbib-sort]{my-jheppub}

\usepackage{amsmath, amssymb}

\usepackage{mathpazo}
\usepackage{bm}
\usepackage{environ}
\usepackage{mathrsfs}
\usepackage{array,arydshln}
\usepackage{booktabs}

\usepackage{graphicx,epsfig}
\usepackage{epic}

\usepackage{tensor} 							
\usepackage{youngtab}

\usepackage{float}
\usepackage[dvipsnames]{xcolor}
\definecolor{maroon}{rgb}{0.8,0.3,0.}

\usepackage{slashed}
\usepackage[nodayofweek]{date time}

\ifShowKeys \usepackage[notcite]{showkeys} \fi

\usepackage{hyperref}

\usepackage{aurical}
\usepackage[T1]{fontenc}

\usepackage{framed}
\definecolor{shadecolor}{RGB}{255, 230, 204}

\allowdisplaybreaks

\newmuskip\pFqmuskip

\newcommand*\pFq[6][8]{%
  \begingroup 
  \pFqmuskip=#1mu\relax
  \mathcode`\,=\string"8000
  \begingroup\lccode`\~=`\,
  \lowercase{\endgroup\let~}\pFqcomma
  {}_{#2}F_{#3}{\left[\genfrac..{0pt}{}{#4}{#5};#6\right]}%
  \endgroup
}

\newcommand*\pFtildeq[6][8]{%
  \begingroup 
  \pFqmuskip=#1mu\relax
  \mathcode`\,=\string"8000
  \begingroup\lccode`\~=`\,
  \lowercase{\endgroup\let~}\pFqcomma
  {}_{#2}\widetilde{F}_{#3}{\left[\genfrac..{0pt}{}{#4}{#5};#6\right]}%
  \endgroup
}

\newcommand{\pFqcomma}{\mskip\pFqmuskip}

\newcommand{\be}{\begin{equation}}
\newcommand{\ee}{\end{equation}}

\newcommand{\mc}{\mathcal }

\newcommand{\bv}{\begin{verbatim}}
\newcommand{\ev}{\end{verbatim}}

\newcommand{\la}{\label}

\title{Large $N$ expansion of  Wilson loops in the Gross-Witten-Wadia matrix model}

\author[a,c]{Eleonora Alfinito}
\author[b,c]{Matteo Beccaria} 

\abstract{
We study the large $N$ expansion of winding Wilson loops in the off-critical 
regime of the Gross-Witten-Wadia (GWW)
unitary matrix model. These have been  
recently considered in \href{https://arxiv.org/abs/1705.06542}{arXiv:1705.06542}
and computed by numerical methods. We present various analytical algorithms for the precise computation of both the perturbative and instanton corrections to the Wilson loops. 
In the gapped phase of the GWW model 
we present the genus five  expansion of the one-cut resolvent that captures all winding loops.
Then, as a complementary tool, 
we apply the Periwal-Shevitz orthogonal polynomial recursion to the GWW model
coupled to suitable sources and show how it  generates all  higher genus
 corrections to any specific loop with
given winding. 
The method is extended to the treatment of instanton effects including higher order
$1/N$ corrections. Several explicit examples are fully worked out and a general formula 
for the next-to-leading correction at general winding is provided. For the simplest cases, 
our calculation checks exact results from the Schwinger-Dyson equations, but the presented
tools have a wider range of applicability.
}

\affiliation[a]{Dipartimento di Ingegneria dell'Innovazione, 
Universit\`a del Salento, 
Via Arnesano, 73100 Lecce, Italy}
\affiliation[b]{Dipartimento di Matematica e Fisica Ennio De Giorgi,\\
Universit\`a del Salento, Via Arnesano, 73100 Lecce, 
Italy} 
\affiliation[c]{Istituto Nazionale di Fisica Nucleare (INFN), Sezione di Lecce} 

\emailAdd{matteo.beccaria@le.infn.it} 

\begin{document}



\maketitle
\flushbottom


\section{Introduction}

The study of the $1/N$ expansion of matrix models is a topic of  clear broad  relevance 
\cite{brezin1993large,Rossi:1996hs,akemann2011oxford,wang2013application}.
Matrix models 
started 
as  elementary building blocks in the large $N$ analysis of ``vector'' models with $O(N)$ flavour
symmetry \cite{Stanley:1968gx}. Their $1/N$ expansion appeared originally as 
an alternative perturbative approach to the determination
of critical exponents \cite{Ma:1973zu} or the starting point for a planar 't Hooft expansion \cite{tHooft:1973alw}. 
In some specific
applications, the space-time dimensionality does not play a role and calculations are
effectively  reduced to 
matrix models, {\em i.e.} ``single-link'' group integrals. A well-known  old example is 
2d Yang-Mills theory on a lattice  where, in the temporal gauge for the Wilson action,  
the partition function factorizes over the lattice links \cite{Gross:1980he}.
In a more recent setting, reduction to matrix models may be due to superconformal symmetry,
as in the case of supersymmetric circular Wilson loop operators in the
$\mc N = 4$ SYM theory that may be reduced to a Gaussian matrix model,
see {\em e.g.} \cite{Erickson:2000af,Drukker:2000rr} and the $\mc N=2$ extension
in \cite{Pestun:2007rz}. In general, for theories admitting an AdS dual, the matrix model formulation
allows for non-trivial tests of AdS/CFT, as in the case of 
BPS Wilson loops  in higher representations in $\mc N=4$ SYM that have
a dual description in terms of D-branes \cite{Drukker:2005kx,Gomis:2006sb,
Gomis:2006im,Hartnoll:2006is,Okuyama:2006jc,Yamaguchi:2007ps,Drukker:2006zk,
Buchbinder:2014nia,Chen-Lin:2016kkk}.
Other modern applications exploit matrix models as quantum gauge theories in zero dimension 
capturing some aspects of more realistic (or complicated) theories, like random surface models
of quantum gravity or 
non critical  and topological string  on specific backgrounds \cite{Marino:2004eq}.

In this paper we reconsider the large $N$ expansion of the Gross-Witten-Wadia (GWW) 
model \cite{Gross:1980he,Wadia:2012fr}. This is the simplest unitary matrix model 
with Wilson action 
\be
S_{GWW}(U) = N\,g\,\text{tr}(U+U^{\dagger}),
\ee
where $g$ is the large $N$ planar  't Hooft coupling. 
The GWW model 
has a large $N$ third order phase transition at $g_{c}=1$. \footnote{For a renormalization group
approach to the matrix model large $N$ transition, see for instance 
\cite{Brezin:1992yc,Higuchi:1994dv}.
} The critical point is captured by a double 
scaling limit \cite{Periwal:1990gf,Periwal:1990qb} and is associated with  type 0B
theory in $d=0$ dimension, {\em i.e.} pure 2-d supergravity \cite{Klebanov:2003wg}. It 
also plays an important role in the description of the adjoint unitary model 
which is used to discuss the non-perturbative aspects of the Hagedorn transition for tensionless 
IIB string theory in AdS \cite{Liu:2004vy,AlvarezGaume:2005fv}. The transition in this 
context is holographically dual to the Hawking-Page transition on the bulk gravity side 
\cite{Witten:1998zw}.

The analysis of the GWW model away from the critical point is particularly interesting 
in the study of its non-perturbative multi-instanton corrections. These have been discussed in 
 \cite{Marino:2008ya} from the point of view of resurgence and trans-series analysis. Multi-instanton configurations in the GWW model may have an  interpretation in terms of eigenvalue tunneling 
 \cite{shenker1991strength,David:1990sk}, and admit a more general characterization
 as complex saddles of the partition function 
 where the $U$ eigenvalues real phases are continued to the complex plane and allowed to 
 accumulate on non-trivial cuts off the unit circle \cite{Buividovich:2015oju,Alvarez:2016rmo}.
 
 The GWW model has a gapless phase for $g<g_{c}$ where  the unitary matrix 
 eigenvalue density is supported on the whole circle. In this phase, 
 the (perturbative) higher genus corrections to the free energy 
 vanish beyond genus zero, while non-perturbative
 corrections remains non trivial. Beyond the third order transition point, for $g>g_{c}$, the GWW model
 is in a different phase where a gap opens in the the eigenvalue density which is non zero 
 on a coupling dependent interval.
 In this phase, the free energy has both perturbative and non-perturbative corrections and 
 these are linked by resurgence \cite{Marino:2008ya}.
 
 Recently, the winding Wilson loops 
 \be
 W_{k}=\frac{1}{N}\,\langle \text{tr}(U^{k})\rangle,
 \ee
  have been analyzed in 
 \cite{Okuyama:2017pil} in both phases of the model. Series expansions in $1/N$ have been proposed 
 for both the purely perturbative genus expansion as well as for the instanton parts of $W_{k}$. \footnote{
 The instanton corrections have an exponentially suppressed pre-factor 
 $\mc {C}\,\exp(-\ell\,N\,S(g))\,$ where $S(g)$ is the instanton action and $\ell$ is the instanton order. 
 As discussed later, this is multiplied by a power series of $1/N$ corrections 
 $\mc W^{(\ell)}_{k,0}(g)+1/N\,\mc W^{(\ell)}_{k,1}(g)+1/N^{2}\,\mc W^{(\ell)}_{k,2}(g)+\dots$ corrections, where $\mc W^{(\ell)}_{k,n}$ are non-trivial functions of the coupling.}
 Technically, these expansions have been 
 obtained by a careful analysis of the numerical exact expressions of $W_{k}$ at finite $N$ and 
 coupling. A  numerical fitting procedure provided the coefficients of the $1/N$ expansion as rational
 functions of the coupling at  genus three accuracy. 
 
 In this paper, we discuss the analytical calculation of these expansions for the winding Wilson loops.
 We first recall some exact results that are consequences of the loop equations of the model. These
 are an important benchmark for the proposed methods whose range of applicability goes
 beyond the Wilson action. We then  focus on the (perturbative) genus expansion in the gapped phase. 
 To this aim, we follow three different approaches. First, 
 we compute the higher genus one-cut resolvent according to the methods
 developed in \cite{Ambjorn:1992gw} for the hermitian matrix models
 and exploiting the results of \cite{Mizoguchi:2004ne} to map the resolvent to the GWW model.
 A systematic calculation provides explicit results at 
 genus five. This allows to write down the genus expansion of all $W_{k}$ at order $1/N^{10}$. 
 The calculation is somewhat straightforward and can be used to provide useful checks of other 
 approaches. 
 
 As a second technique, we compute $W_{k}$ by coupling the GWW model 
 to  auxiliary sources $\bm\rho$
  \be
  \la{y1}
S_{GWW}(U; \bm{\rho}) = N\,g\,\text{tr}(U+U^{\dagger})+\sum_{k=2}^{\infty}
\rho_{k}\,\text{tr}(U^{k}+(U^{\dagger})^{k}).
 \ee
 Derivatives of the free energy with respect
 to the sources $\rho_{k}$ compute  $W_{k}$. The free energy associated with the extended action
 (\ref{y1}) is computed by the
 orthogonal polynomial methods developed for hermitian matrix models in 
 \cite{Bessis:1979is,Bessis:1980ss,DiFrancesco:1993cyw}
 and extended to unitary matrix models in  \cite{Goldschmidt:1979hq,Periwal:1990qb,Periwal:1990gf}.
 Each source $\rho_{k}$ is associated with an increasingly involved recursion relation for the 
 orthogonal polynomial coefficients. This can be solved perturbatively at large $N$. We clarify some
 technical aspect of the procedure and perfectly reproduce the results for $W_{k}$ obtained 
 by the resolvent approach. These calculations extend the results of \cite{Okuyama:2017pil}
 and confirm them up to some discrepancies that are important for reconciliation with 
 old exact results relating $W_{1}$ and $W_{2}$ at finite 
 coupling and $N$.
 
 Finally, we discuss a third (practical) approach that is based on some peculiarities of the large coupling 
 expansion of the finite $N$ expressions of $W_{k}$. Guided by an educated guess for the structure of the
 genus expansion we can provide, with very modest effort, quite long genus expansions for the 
 winding Wilson loops. This is somewhat interesting because it is cumbersome to extend
 the resolvent method at high orders in $1/N$. This is not a difficulty for the
 orthogonal polynomial method, but in that case one has to deal with a recursion relation whose
 complexity increases with $k$. Instead, the proposed  analytic bootstrap of finite $N$ data can treat
 with minor effort with higher $k$ and even for more complicated observables
 like Wilson loops in general small representations. 
 
 As a final result, according to the ideas of \cite{Marino:2008ya}, we also discuss the use of the 
 orthogonal polynomial method for the GWW model coupled to sources
 to compute the non perturbative instanton corrections
 to the winding Wilson loop. In particular, we clarify various technical issues and 
 provide  explicit examples in the gapless phase. 
 
 The plan of the paper is the following. In Sec.~(\ref{sec:intro}), we briefly recall some basic
 facts about the GWW model and its large $N$ expansion. 
Sec.~(\ref{sec:gapped}) is devoted to the analysis of the high genus expansion in the gapless phase.
In more details, in Sec.~(\ref{sec:resolvent}) we present the one-cut resolvent at genus five.
In Sec.~(\ref{sec:PS}), we discuss how to apply the Periwal-Shevitz recursion method to the 
GWW model coupled to suitable sources in order to compute $W_{k}$. In 
 Sec.~(\ref{sec:boot}), we reproduce and extend the previous results by an analytic bootstrap
 procedure that exploits a simple combination of educated insight and (small) finite $N$ data. 
 Finally,  in Sec.~(\ref{sec:instanton}), the methods presented   in Sec.~(\ref{sec:PS}) 
 are used for the  precise calculation of large $N$ effects in the 
 instanton contribution to $W_{k}$ in the gapless phase. 

\section{The Gross-Witten-Wadia model}
\la{sec:intro}

The 
partition function of the GWW model, at finite matrix dimension $N$ and coupling $g$, admits the following
exact expression 
 \cite{Wadia:2012fr,Bars:1979xb}\footnote{We shall follow the notation of 
 \cite{Okuyama:2017pil} to simplify comparison.}
\be
\la{2.1}
Z(N, g) = \int_{U(N)}dU\,\exp\bigg[\frac{N\,g}{2}\,\text{tr}(U+U^{\dagger})\bigg] = 
\det\left[I_{n-m}(N\,g)\right]_{n,m=1,\dots, N},
\ee
where $I_{\nu}(x)$ is the modified Bessel function of the first kind.
The observables we are going to study are the winding Wilson loops whose definition and 
exact expression read
\be
\la{2.2}
W_{k}(N,g) = \frac{1}{N}\,\langle \text{tr}(U^{k})\rangle = \frac{1}{N}\,\text{tr}(M_{0}^{-1}\,M_{k}), \qquad 
(M_{k})_{n,m} = I_{k+n-m}(N\,g).
\ee
We remark that the  (standard) Wilson loop $W_{1}$ can be computed directly from the free energy
(\ref{2.1}) according to the obvious relation
\be
\la{2.3}
W_{1}(N,g) = \frac{1}{N^{2}}\partial_{g}\,\log Z(N, g).
\ee
In particular, from the known expansion of the free energy, this implies that 
\be
\la{2.4}
\lim_{N\to \infty} W_{1}(N,g) = \begin{cases}
\frac{g}{2}, & g<1, \\
1-\frac{1}{2\,g}, & g>1.
\end{cases}
\ee
At fixed $g$ and large $N$, the winding Wilson loops can be computed in terms of the 
asymptotic distribution $\rho(\theta)$ of the eigenvalues $e^{i\theta}$ of the $U(N)$ matrix \cite{Gross:1980he}. This gives the result
\be
\la{2.5}
\lim_{N\to \infty} W_{k}(N,g) = \begin{cases}
0, & g<1, \\
\frac{1}{k-1}\left(1-\frac{1}{g}\right)^{2}\,P_{k-2}^{(1,2)}\left(1-\frac{2}{g}\right), & g>1,
\end{cases}
\ee
where $P^{(a,b)}_{n}(z)$ are Jacobi polynomials. Notice that the second derivative
of $W_{k}$ has a finite jump at the transition point $g=1$. The point $g=1$ separates
an ungapped phase $(g<1)$ where the large $N$ eigenvalue distribution $\rho(\theta)$
is supported on the whole circle,
and a gapped phase $(g>1)$ where a gap opens in the distribution $\rho(\theta)$.

\subsection{Loop equations for winding Wilson loops}

An alternative approach to the large $N$ limit is based on the relation between the GWW model and 
Yang-Mills theory in two dimensions  in the lattice Wilson formulation. One can show that 
the  higher winding Wilson loops obey an algebraic equation at fixed $g$ and large $N$. 
This follows from the Migdal-Makeenko loop equation \cite{Makeenko:1979pb,Migdal:1984gj}
that,
after using large $N$ factorization,  leads to the following generating function 
for $w_{n}\equiv W_{n}(N\to \infty)$ \cite{Paffuti:1980cs}
\begin{align}
\la{2.6}
\sum_{n=0}^{\infty}w_{n}\,x^{n} &= 
\frac{g}{4x}\, \sqrt{\left(\frac{2 x}{g}+x^2+1\right)^2-\frac{4 x^2 (g-2 w_{1})}{g}}
+\frac{g \left(x^2-1\right)+2 x}{4 x}\notag \\
&= 1+w_{1}\,x+\left(
1-\frac{2\,w_{1}}{g}\right)\,x^{2}+\left[
-\frac{2}{g}+w_{1}\,\left(1+\frac{4}{g^{2}}\right)-\frac{2\,w_{1}^{2}}{g}
\right]\,x^{3}+\dots\, .
\end{align}
Starting with (\ref{2.4}), the relations (\ref{2.6}) are equivalent to (\ref{2.5}). However, it is 
useful to remark that the relation linking $w_{2}$ to $w_{1}$ is special being exact even at  finite $N$
 \cite{Wadia:1980rb,Friedan:1980tu} \footnote{For a given $N$, this can be easily 
 checked explicitly from the exact expressions in (\ref{2.2}). }
\be
\la{2.7}
W_{2}(N,g) = 1-\frac{2}{g}\,W_{1}(N,g),
\ee
and is thus an important constraint on the computations. Actually, the exact relation (\ref{2.7})
may be generalized to higher winding loops by inspecting the loop equations at specific low
windings. Using, for instance, the results of \cite{Green:1980bg} it is easy to prove the next two
exact relations ($W_{k} \equiv W_{k}(N,g)$)
\begin{align}
\la{kuz1}
W_{3} &= -\frac{2}{g}+\left(1+\frac{4}{g^{2}}+\frac{2}{g^{2}\,N^{2}}\right)\,W_{1}
-\frac{2}{g}\,W_{1}^{2}-\frac{2}{g\,N^{2}}\,\partial_{g}W_{1},  \notag \\
W_{4} &= 1+\frac{4}{g^{2}}+\frac{8}{g^{2}N^{2}}+\left(
-\frac{8}{g}-\frac{8}{g^{3}}-\frac{28}{g^{3}N^{2}}\right)\,W_{1}
+\frac{12}{g^{2}}\,W_{1}^{2}+\frac{12}{g^{2}N^{2}}\,\partial_{g}W_{1},
\end{align}
that are additional constraints holding at finite $N$ and $g$. Going at higher winding involves 
more complicated correlators. For instance, the loop equation for $W_{5}$ links it to 
$\langle \text{tr}(U^{2})^{2}\rangle$. Actually, as we explained in the Introduction, we are going to 
use results like (\ref{2.7}) and (\ref{kuz1}) as a test in the GWW model of more general tools
that, in principle, may be used for unitary models with actions different from the simplest Wilson one.

\subsection{Structure of $1/N$ and non-perturbative corrections}

The free energy of the GWW model admits the large $N$ genus expansion 
\begin{align}
\la{2.8}
F(N,g) &= \log Z(N,g) = F^{\rm pert}(N, g)+F^{\rm inst}(N, g), \notag \\
F^{\rm pert}(N, g)&=\sum_{\ell\ge 0}N^{2-2\,\ell}\,F_{\ell}(g),
\end{align}
where $F^{\rm inst}(N, g)$ are instanton contributions exponentially suppressed at large $N$. 
The genus expansion coefficients $F_{\ell}(g)$ are trivial in the ungapped phase
\be
F_{0}(g) = \frac{g^{2}}{2},\qquad F_{\ell\ge 1}(g)=0.
\ee
Instead, in the gapped phase, all coefficients $F_{\ell}(g)$ are non zero. They start 
with
\be
\la{2.10}
F_{0}(g) = g-\frac{1}{2}\,\log g-\frac{3}{4}, \quad
F_{1}(g) = \zeta'(-1)-\frac{1}{12}\,\log N-\frac{1}{8}\log(1-1/g),
\ee
and, for $\ell\ge 2$, may be written in the general form  
\be
F_{\ell\ge 2}(g) = \frac{B_{2\,\ell}}{2\,\ell\,(2\,\ell-2)} +\frac{1}{(g-1)^{3\,\ell-3}}\sum_{n=0}^{\ell-2}
c_{n}^{(\ell)}\,g^{n},
\ee
where the coefficients $c_{n}^{(\ell)}$ may be systematically  
computed at higher genus by solving the {\em pre-string} equation
governing the partition function 
 \cite{Goldschmidt:1979hq,Periwal:1990qb,Periwal:1990gf}
 as discussed in details in \cite{Marino:2008ya}. 
 The non perturbative part $F^{\rm inst}(g)$ in (\ref{2.8}) is non trivial in both phases. Its detailed structure
 as well as the relation with resummation of the genus expansions have been fully elucidated in 
  \cite{Marino:2008ya}.
  
In the case of winding Wilson loops, the recent results of \cite{Okuyama:2017pil} suggest 
that  we can write again a decomposition similar to (\ref{2.8})
\begin{align}
\la{2.12}
W_{k}(N,g) &= W_{k}^{\rm pert}(N,g)+W_{k}^{\rm inst}(g),\notag \\
W_{k}^{\rm pert}(N,g) &= \sum_{\ell\ge 0}N^{2-2\ell}W_{k, \ell}(g),
\end{align}
where the genus expansion correcting (\ref{2.4}) and (\ref{2.5}) is non trivial in the gapped phase. 
The instanton contribution is present in both phases and takes the general form 
\be
\la{2.13}
W_{k}^{\rm inst}(N,g) = \sum_{\ell=2}^{\infty}\frac{e^{-\ell\,N\,S(g)}}{4\,\pi\,N^{2}}\sum_{n=0}^{\infty}\frac{\mc W^{(\ell)}_{k,n}(g)}{N^{n}},
\ee 
where the instanton action is 
\be
\la{2.14}
S(g) = \cosh^{-1}(1/g)-\sqrt{1-g^{2}},
\ee
and the exponential prefactor comes together with 
an infinite perturbative tail factor associated with the coefficients $\mc W_{k,n}^{(\ell)}$.

In the following sections, we shall first focus on the gapped phase and discuss how to derive 
analytically the perturbative contributions $W_{k, \ell}(g)$ in (\ref{2.12}) at high genus. Later,
we shall address the computation of the coefficients $\mc W_{k,n}^{(\ell)}$ in the non-perturbative
contribution focusing on the ungapped case where this is the only non trivial correction 
to the winding Wilson loops at large $N$.
 
 \section{High genus expansion in the gapped phase}
\la{sec:gapped}

In order to clarify the motivations of our analysis, let us start by remarking that 
from the relation (\ref{2.3}) and the known expansion of the free energy, we obtain the exact expansion of 
  $W_{1}^{\rm pert}$ in the gapped phase
 \be
 \la{3.1}
 W_{1}^{\rm pert} = 1-\frac{1}{2\,g}-\frac{1}{N^{2}}\,\frac{1}{8\,(g-1)\,g}-\frac{1}{N^{4}}\,
 \frac{9}{128\,(g-1)^{4}}-\frac{1}{N^{6}}\,\frac{9\,(25\,g+17)}{1024\,(g-1)^{7}}+\dots.
 \ee
 For the winding loops, the results of \cite{Okuyama:2017pil} -- denoted below by a tilde -- are based on a numerical fitting
 procedure and in the case of $W_{2}$ 
and $W_{3}$ they read
\begin{align}
\la{3.2}
\widetilde{W}_{2}^{\rm pert} &= \frac{(g-1)^{2}}{g^{2}}+\frac{1}{N^{2}}\,\frac{1}{4\,(g-1)\,g^{2}}
+\frac{1}{N^{4}}\,\frac{9}{64\,(g-1)^{4}\,g}+\frac{1}{N^{6}}\,
\frac{451\,g^{2}+297\,g+23}{1024\,(g-1)^{7}\,g^{2}}+\dots, \notag \\
\widetilde{W}_{3}^{\rm pert} &= \frac{(g-1)^{2}(2\,g-5)}{2\,g^{3}}
+\frac{1}{N^{2}}\,\frac{10-28\,g+15\,g^{2}}{8\,(g-1)\,g^{3}}
+\frac{1}{N^{4}}\,\frac{3\,(20-90\,g+96\,g^{2}-35\,g^{3})}{128\,(g-1)^{4}\,g^{3}}
+\dots.
\end{align}
Similar expansions
at genus two are also provided for the higher winding Wilson loops.  
The expression $\widetilde{W}_{2}$
is not compatible with the exact relation (\ref{2.7}). Indeed, that relation implies, using 
the expansion (\ref{3.1}), the result
\be
W_{2}^{\rm pert} = \frac{(g-1)^{2}}{g^{2}}+\frac{1}{N^{2}}\,\frac{1}{4\,(g-1)\,g^{2}}
+\frac{1}{N^{4}}\,\frac{9}{64\,(g-1)^{4}\,g}+\frac{1}{N^{6}}\,
\frac{9\,(17+25\,g)}{512\,(g-1)^{7}\,g}+\dots,
\ee
that differs from the first of (\ref{3.2}) at the $N^{-6}$ level.
For this reason, it seems important to compute analytically the expansions of $W_{k}^{\rm pert}$ in order
to test the numerical fitting in \cite{Okuyama:2017pil}. An analysis of the instanton corrections 
in the gapped phase will be addressed later in Sec.~(\ref{sec:instanton}).

\subsection{Genus expansion of the one-cut resolvent}
\la{sec:resolvent}

The solution of the loop equations for all winding loops  $W_{k}$ at a given genus order may be 
achieved by resolvent techniques.
The GWW model can be mapped to a non-polynomial hermitian matrix model 
according to the results of  \cite{Mizoguchi:2004ne}. In particular, the gapped phase of the GWW model 
is associated with a (symmetric) one-cut resolvent. 
For a general hermitian matrix model, there exists a systematic iterative scheme for the resolvent
calculation with explicit results up to genus two \cite{Ambjorn:1992gw}.\footnote{
Further discussion and details 
 can be found in \cite{Akemann:2001st,Marino:2007te}.
 }
We briefly discuss the method
 and then apply to the GWW model at genus five. 
The resolvent of the GWW model is defined by the expectation value
\be
G(t) = \frac{i}{N}\,\left\langle\text{tr}\frac{t+U}{t-U}\right\rangle,
\ee
and the values of $W_{k}$ can be read from its large $t$ expansion. We shall exploit the relation 
\be
\la{3.5}
G(t) = (1+z^{2})\,\omega(z)-z, \qquad t = \frac{1+i\,z}{1-i\,z},
\ee
where $\omega(z)$ is the resolvent of a hermitian matrix model in terms of $M=M^{\dagger}$
\be
\omega(z) = \frac{1}{N}\left\langle\text{tr}\frac{1}{z-M}\right\rangle,\qquad 
U = \frac{i-M}{i+M}.
\ee
The relation between the GWW  partition function and that of the hermitian matrix model is 
simply
\begin{align}
& \int dU\, e^{-N\,\text{tr}W(U)} = \int dM\, e^{-N\,\text{tr} V(M)}, \qquad
 V(z) = W(t)+\log(1+z^{2}).
\end{align}
According to \cite{Ambjorn:1992gw}, we also introduce  the 
2-point (connected) resolvent
\be
\la{3.8}
\omega(z, z') = \left\langle\text{tr}\frac{1}{z-M}\,\frac{1}{z'-M}\right\rangle.
\ee
Both $\omega(z)$ and $\omega(z,z')$ admit a genus expansion 
\be
\la{3.9}
\omega = \sum_{g\ge 0}N^{-2\,g}\,\omega_{g}.
\ee
The first Makeenko loop equation for the hermitian matrix model reads \cite{Makeenko:1991tb}
\be
\la{3.10}
\oint \frac{d\zeta}{2\,\pi\,i}\frac{V'(\zeta)}{z-\zeta}\,\omega(\zeta) = \omega(z)^{2}+\frac{1}{N^{2}}\,
\omega(z,z),
\ee
where the integral is along a simple positive curve enclosing all singularities of $\omega(\zeta)$.
The genus expansion of (\ref{3.10}) reads 
\be
\la{3.11}
\widehat{K}\,\omega_{g}(z) = \sum_{g'=1}^{g-1}\omega_{g'}(z)\,\omega_{g-g'}(z)
+\frac{\delta\omega_{g-1}(z)}{\delta V},
\ee
where $\widehat{K}$ is a suitable integral linear operator and the loop insertion derivative
$\delta/\delta V$ has a complicated but explicit expression that can be found in \cite{Ambjorn:1992gw}.
 The iterative solution of (\ref{3.11}) starts by expressing the resolvent as 
 \be
 \la{3.12}
 \omega_{g}(z) = \sum_{n=1}^{3g-1}\bigg[A_{g}^{+(n)}\,\chi^{+(n)}(z)
 +A_{g}^{-(n)}\,\chi^{-(n)}(z)\bigg],
 \ee
 where the functions $\chi^{\pm (n)}(z)$  are explicit eigenfunctions of $\widehat K$
 that can be given explicitly and  depend on the potential $V$
 through the so-called moments ($k\ge 1$)
\be
M_{k} = \oint\frac{d\zeta}{2\pi\,i}\frac{V'(\zeta)}{(\zeta-x)^{k+1/2}(\zeta-y)^{1/2}}, \quad
J_{k} = \oint\frac{d\zeta}{2\pi\,i}\frac{V'(\zeta)}{(\zeta-x)^{1/2}(\zeta-y)^{k+1/2}},
\ee
where $x$, $y$ are the real endpoints of the resolvent cut $[y,x]\subset\mathbb R$.
Plugging (\ref{3.12}) in (\ref{3.11}), we can 
determine the constants $A_{g}^{\pm(n)}$ and thus the resolvent. The lowest order results 
are 
\begin{align}
\la{3.14}
\omega_{0}(z) &= \frac{A^2 \,z\,\left(2\, g+1+z^2\right)-2 \sqrt{A^2+1} \sqrt{z^2-A^2}}{A^2 \left(z^2+1\right)^2}, \notag \\
\omega_{1}(z) &=
-\frac{A^4 \sqrt{A^2+1} \left(A^2-2 z^2-1\right)}{16 \left(z^2-A^2\right)^{5/2}}, \notag \\
\omega_{2}(z) &= \frac{A^{8}\,\sqrt{1+A^{2}}}{1024\,(z^{2}-A^{2})^{11/2}}\,\bigg[
36 A^{10}-180 A^8 z^2+9 A^6 \left(41 z^4+2 z^2+1\right)\notag \\
&-2 A^4 \left(206 z^6+57 z^4+36 z^2+5\right)+A^2
   \left(292 z^8+300 z^6+285 z^4+118 z^2+21\right)\notag \\
   & +12 z^2 \left(18 z^6+34 z^4+26 z^2+7\right)
\bigg],
\end{align}
 where $A = \frac{1}{\sqrt{g-1}}$ 
is the endpoint of the (symmetric) cut $[y,x]\equiv [-A,A]$. Inserting in (\ref{3.5})  
the genus expansion of the resolvent, see (\ref{3.9}), and expanding at large $t$, one obtains the genus expansion 
of all winding Wilson loops. We pushed the calculation extending the list in (\ref{3.14})
up to genus 5. \footnote{The rather unwieldy expressions of the coefficients in (\ref{3.12}) and of the
resolvent are available under request.}
The standard Wilson loop $W_{1}$ reads
\begin{align}
\la{3.15}
W_{1}^{\rm pert} &= 1-\frac{1}{2\,g}-\frac{1}{N^{2}}\,\frac{1}{8\,(g-1)\,g}-\frac{1}{N^{4}}\,
 \frac{9}{128\,(g-1)^{4}}\notag \\
 & -\frac{1}{N^{6}}\,\frac{9\,(25\,g+17)}{1024\,(g-1)^{7}}
 -\frac{1}{N^{8}}\frac{9 \left(6125\, g^2+10750\, g+2381\right)}{32768\, (g-1)^{10}}\notag \\
 &
 -\frac{1}{N^{10}}\,\frac{9 \left(694575\, g^3+2160925\, g^2+1293325\, g+140927\right)}
 {262144 (g-1)^{13}}+\dots.
 \end{align}
This expression agrees with (\ref{3.1}) and we checked agreement of higher genus corrections with the 
pre-string expansion of the free energy. The resolvent prediction for the second Wilson loop is 
 \begin{align}
W_{2}^{\rm pert} &= \frac{(g-1)^{2}}{g^{2}}
+\frac{1}{N^{2}}\,\frac{1}{4\, (g-1)\, g^2}
+\frac{1}{N^{4}}\,\frac{9}{64 \,(g-1)^4\, g}\notag \\
&+\frac{1}{N^{6}}\,\frac{9 (25\,   g+17)}{512\, (g-1)^7\, g}
+\frac{1}{N^{8}}\,\frac{9 \left(6125\, g^2+10750\, g+2381\right)}{16384 \,(g-1)^{10} \,g}\notag \\
&
+\frac{1}{N^{10}}\,\frac{9 \left(694575\, g^3+2160925\, g^2+1293325\, g+140927\right)}
{131072\, (g-1)^{13}\, g}+\dots.
\end{align}
The exact relation (\ref{2.7}) is satisfied, as it should. The next Wilson loop is \footnote{
It is instructive to insert this expansion into the large $N$ relation (\ref{2.6}). As expected, the $1/N^{2}$
terms do not cancel due to non-factorization corrections to the loop equations.
}
 \begin{align}
 \la{3.17}
W_{3}^{\rm pert} &=\frac{(g-1)^2 (2 g-5)}{2 g^3}+\frac{1}{N^{2}}\,\frac{15 g^2-28 g+10}{8 (g-1) g^3}
-\frac{1}{N^{4}}\,\frac{3 \left(35 g^3-96
   g^2+90 g-20\right)}{128 \,(g-1)^4 g^3 }\notag \\
   & -\frac{1}{N^{6}}\,\frac{27 \left(35 g^3-85 g^2+62
   g+30\right)}{1024 (g-1)^7 g^2}\notag \\
   & -\frac{1}{N^{8}}\,\frac{9 \left(17325 g^4-27010 g^3-11765
   g^2+65748 g+13470\right)}{32768 (g-1)^{10} g^2}\notag \\
   & -\frac{1}{N^{10}}\,\frac{9 \left(1576575
   g^5-537075 g^4-5526245 g^3+8822345 g^2+7716666 g+816990\right)}
   {262144 (g-1)^{13}g^2}+\dots.
\end{align}
It is clear that similar results for any $W_{k}$ can be written immediately, at this genus order. 
Examples are \footnote{Notice that the $1/N^{4}$ contribution in 
$W_{5}$ is different from \cite{Okuyama:2017pil}.}
\begin{align}
\la{3.18}
W_{4}^{\rm pert} &= \frac{(g-1)^2 \left(g^2-6 g+7\right)}{g^4}+\frac{1}{N^{2}}
\frac{16 g^3-70 g^2+90 g-35}{2 (g-1) g^4}\notag \\
& +\frac{1}{N^{4}}
\frac{226 g^3-624 g^2+561 g-154}{32 (g-1)^4 g^4}
+\frac{1}{N^{6}}\,\frac{9 \left(202 g^3-470 g^2+275 g+35\right)}
{256 (g-1)^7 g^3}\notag\\
&+\frac{1}{N^{8}}\,\frac{9 \left(32250 g^4-42564 g^3-37885 g^2+58950
   g+8505\right)}{8192 (g-1)^{10} g^3}\notag \\
   & +\frac{1}{N^{10}}\frac{9 \left(2908150
   g^5+30850 g^4-10788813 g^3+5267705 g^2+6286275 g+585585\right)}{65536 (g-1)^{13}
   g^3 }+\dots, \notag \\
W_{5}^{\rm pert} &= \frac{(g-1)^2 \left(2 g^3-21 g^2+56 g-42\right)}{2 g^5}+\frac{1}{N^{2}}
\frac{5 \left(35 g^4-260
   g^3+630 g^2-616 g+210\right)}{8 (g-1) g^5}\notag \\
   & +\frac{1}{N^{4}}\,\frac{6615 g^5-32672
   g^4+60438 g^3-52332 g^2+21098 g-3192}{128 (g-1)^4 g^5}\notag \\
   & -\frac{1}{N^{6}}\frac{45
   \left(693 g^6-3923 g^5+9670 g^4-12250 g^3+7910 g^2-2394 g+336\right)}{1024
   (g-1)^7 g^5 }\notag \\
   & -\frac{1}{N^{8}}\frac{45 \left(33033 g^6-173082
   g^5+416379 g^4-341620 g^3-103320 g^2+179172 g+8694\right)}{32768
   (g-1)^{10} g^4 }\notag \\
   & -\frac{1}{N^{10}}\frac{45}{262144 (g-1)^{13} g^4}\bigg(1756755 g^7-7613879
   g^6+13944571 g^5+15923913 g^4\notag \\
   & -51364240 g^3+14031360 g^2+16537374
   g+1073898\bigg)+\dots\ . 
 \end{align}
 Notice that the expansions for $W_{3}$ and $W_{4}$ are in full agreement with (\ref{kuz1}).
 
 The advantage of the resolvent method is that  the fixed genus corrections to 
 all $W_{k}$ are obtained altogether in one shot. However, pushing the calculation at higher genus 
 becomes more and more 
 cumbersome. Besides, in the ungapped phase, the use of this approach to compute 
 the non-perturbative instanton corrections is quite involved, see for instance  
  \cite{Marino:2007te}.
 In the next section, we discuss a different complementary  strategy. Its  complexity 
 increases with $k$, but -- at fixed $k$ -- it easily  generates 
 the full $1/N$ expansion with minor effort. It will be later extended to capture instanton corrections 
 following the ideas put forward in \cite{Marino:2008ya}.
 
 As a final comment about the methods presented in this section, we remark that 
 the 2-point generating function  (\ref{3.8}) is 
 a byproduct of the  resolvent calculation.
 This may be exploited to 
 get expansions for other observables involving product of traces of powers of $U$, like those
 appearing in the study of Wilson loops in small representations.
 
\subsection{The Periwal-Shevitz recursion for the GWW model coupled to sources}
\la{sec:PS}

Let us change variables and 
introduce the (string) coupling $g_{s} = 1/(N\,g)$ to trade the matrix dimension $N$. We can 
consider the partition function 
\be
\la{3.19}
Z_{V}(N, g_{s}) = \int dU\,\exp\bigg[\frac{1}{g_{s}}\,\text{tr}[V(U)+V(U^{\dagger})]\bigg],
\ee
for a generic potential $V(z)$. The idea is simply to couple the GWW model to suitable sources
in order to generalize the relation (\ref{2.3}) to higher winding. To this aim
we consider  
\be
\la{3.20}
V(U) = \frac{1}{2}\,\bigg(U+\rho\,g_{s}\,U^{k}\bigg),
\ee
and extract the winding Wilson loop $W_{k}$ from 
\be
W_{k} = \frac{1}{N}\,\lim_{\rho\to 0}\partial_{\rho}\log Z_{V}.
\ee
The point is that (\ref{3.19}) may be treated in full generality by using the method of orthogonal polynomials
developed in \cite{Bessis:1979is,Bessis:1980ss} and extended to unitary models in 
\cite{Periwal:1990gf}. To remind the basic facts, 
one introduces monic polynomials $p_{n}(z) = z^{n}+\cdots$ that are 
orthogonal with respect to the circle measure $d\mu$
\begin{align}
\la{3.22}
\oint d\mu\,p_{n}(z)\,p_{m}(1/z) = h_{n}\,\delta_{nm}, \qquad
 d\mu = \frac{1}{2\,\pi\,i}\frac{dz}{z}\,\exp\bigg[\frac{1}{g_{s}}(V(z)+V(1/z))\bigg].
\end{align}
Quite generally, it can be shown that the polynomials $p_{n}(z)$ obey the functional recursion
\be
p_{n+1}(z) = z\,p_{n}(z)+f_{n}\,z^{n}\,p_{n}(1/z), \qquad \frac{h_{n+1}}{h_{n}} = 1-f_{n}^{2}.
\ee
The normalizations $h_{n}$, {\em cf.} (\ref{3.22}), and their ratios $r_{n}=h_{n}/h_{n-1}$, determine
 the exact partition function in (\ref{3.19})
 \be
 Z_{V} = \prod_{n=0}^{N-1}h_{n} = h_{0}^{N}\,\prod_{n=1}^{N}r_{n}^{N-n}.
 \ee
It is convenient to introduce $t=1/g = N\,g_{s}$ and write the free energy as \footnote{In some cases,
it may be  
convenient to subtract a reference free energy. Here, we are interested in the linear in $\rho$
part of $F$ and this subtraction is not needed.}
\be
\la{3.25}
g_{s}^{2}\,F = \frac{t^{2}}{N}\,\log h_{0}+\frac{t^{2}}{N}\,\sum_{n=1}^{N}
\bigg(1-\frac{n}{N}\bigg)\,\log r_{n}.
\ee
We remark that the quantity $h_{0} = \oint d\mu$ is a non-trivial function of $\rho$. The other
relevant quantities $r_{n}$ obey a recursion that depends on $V$. Defining for convenience
$s_{n}^{2} = 1-f_{n}^{2} = r_{n+1}$, one can show that the following identity holds
\be
\la{3.25}
g_{s}\,(n+1)\,\frac{f_{n}^{2}}{s_{n}} = \oint d\mu\,p_{n+1}(z)\,p_{n}(1/z)\,\frac{d}{dz}\bigg[
V(z)+V(1/z)\bigg].
\ee
As shown in \cite{Periwal:1990qb}, the {\em r.h.s.} of (\ref{3.25}) may be evaluated 
in explicit form by means of the  Migdal-Gross operatorial formalism 
\cite{Gross:1989vs,Gross:1989aw}. The result is the remarkable formula
\be
\la{3.27}
g_{s}\,(n+1)\,\frac{f_{n}^{2}}{s_{n}} = \oint \frac{du}{2\,\pi\,i}\,\mathcal{S}\bigg[
V'(z_{n+1})-\frac{1}{z_{n+1}^{2}}\,V'(1/z_{n+1})
\bigg],
\ee
where, in the square bracket, we replace the non-commutative formal expansions
\begin{align}
z_{m} &= u\star s_{m}-f_{m-1}\star f_{m}-u^{-1}\star f_{m-2}\star s_{m-1}\star f_{m}
-u^{-2}\star f_{m-3}\star s_{m-2}\star s_{m-1}\star f_{m}- \cdots, \notag \\
z_{m}^{-1} &= u^{-1}\star s_{m-1}-f_{m}\star f_{m-1}-u\star f_{m+1}\star s_{m}\star f_{m-1}
-u^{2}\star f_{m+2}\star s_{m+1}\star s_{m}\star f_{m-1}-\cdots\ , 
\end{align}
and the $\mathcal S$ operation brings the powers  $u^{k}$ to the left using the commutation
 relations 
\be
X\star s_{m}\star u^{k}\star Y = X\star u^{k}\star s_{m+k}\star Y ,\qquad  
X\star f_{m}\star u^{k}\star Y = X\star u^{k}\star f_{m+k}\star Y.
\ee
After this procedure, we simply pick the residue in $u$ and make commutative the $\star$-product. 
For the potentials $V(z) = z^{k}/2$ with $k=1,2$
 we obtain the known recursions, see \cite{Periwal:1990gf},
\begin{align}
\la{3.30}
k&=1, \quad g_{s}\,(n+1)\,f_{n} = \frac{1}{2} (1-f_{n}^{2})\,(f_{n-1}+f_{n+1}),\notag \\
k&=2, \quad g_{s}\,(n+1)\,f_{n} =  (1-f_{n}^{2})\,(
-f_{n-2} f_{n-1}^2-f_n f_{n-1}^2-2 f_n f_{n+1} f_{n-1}-f_n
   f_{n+1}^2\notag \\
   & +f_{n-2}-f_{n+1}^2 f_{n+2}+f_{n+2}
).
\end{align}
Higher $k$ may be worked out algorithmically according to the above rules. 
For instance, for $k=3$, relevant to the 
computation of $W_{3}$, we get the 26-term recursion
\begin{align}
\la{3.31}
k&=3, \quad g_{s}\,(n+1)\,f_{n} = \frac{3}{2} (1-f_{n}^{2})\,
(f_{n-2}^2 f_{n-1}^3+f_n^2 f_{n-1}^3+2 f_{n-2} f_n
   f_{n-1}^3+f_{n-3} f_{n-2}^2 f_{n-1}^2\notag \\
   & -f_{n-3} f_{n-1}^2+3 f_n^2 f_{n+1}
   f_{n-1}^2+2 f_{n-2} f_n f_{n+1} f_{n-1}^2-f_{n+1} f_{n-1}^2-f_{n-2}^2 f_{n-1}\notag \\
   & +3
   f_n^2 f_{n+1}^2 f_{n-1}-f_{n+1}^2 f_{n-1}-2 f_{n-2} f_n f_{n-1}+2 f_n f_{n+1}^2
   f_{n+2} f_{n-1}-2 f_n f_{n+2} f_{n-1}\notag \\
   & +f_n^2 f_{n+1}^3-f_{n-3}
   f_{n-2}^2+f_{n+1}^3 f_{n+2}^2-f_{n+1} f_{n+2}^2+f_{n-3}-2 f_{n-2} f_n f_{n+1}+2
   f_n f_{n+1}^3 f_{n+2}\notag \\
   & -2 f_n f_{n+1} f_{n+2}-f_{n+1}^2 f_{n+3}+f_{n+1}^2
   f_{n+2}^2 f_{n+3}-f_{n+2}^2 f_{n+3}+f_{n+3}).
\end{align}
The longer $k=4$ case is reported in App.~(\ref{app:PS4}).

\subsubsection{Continuum limit and large $N$ expansion}

According to  \cite{Marino:2008ya}, whose notation we adopt, we can make the following 
replacements at large $N$ 
\be
g_{s}\,n \to z,\quad r_{n-1}\to R(z, g_{s}).
\ee
For instance, the recursion for $k=1$ becomes the functional equation, {\em cf.} Eq.~(4.18) of 
\cite{Marino:2008ya},
\be
\la{3.33}
z\,\sqrt{1-R(z,g_{s})} = \frac{1}{2}\,R(z, g_{s})\,\bigg[
\sqrt{1-R(z+g_{s}, g_{s})}+\sqrt{1-R(z-g_{s}, g_{s})}\bigg].
\ee
Higher $k$ recursion as in (\ref{3.30}) and (\ref{3.31}) are treated in the same way and lead
to structurally similar equations.
An important remark is that in the potential (\ref{3.20}) the additional $\rho$-dependent source
term comes with an explicit $g_{s}$ factor making a perturbative treatment (in 
both $g_{s}$ and $\rho$) feasible.

The large $N$ expansion of the free energy is achieved by standard methods, see the clean
exposition in  \cite{Marino:2008ya}. We can rewrite (\ref{3.25}) in the form 
\begin{align}
\la{3.34}
g_{s}^{2}\,F &= \int_{0}^{t}dz\,(t-z)\,\log R(z,g_{s})+\sum_{p=1}^{\infty}g_{s}^{2p}\,\frac{B_{2p}}
{(2p)!}\frac{d^{2p-1}}{dz^{2p-1}}\left. \bigg[(t-z)\,\log R(z,g_{s})\bigg]\right|_{z=0}^{z=t}\notag \\
&+\frac{t\,g_{s}}{2}\bigg[2\,\log h_{0}-\log R(0, g_{s})\bigg].
\end{align}
A consequence of (\ref{3.34}) is 
\be
\la{3.35}
g_{s}^{2}\,F''(t) = \log R(t, g_{s})-\sum_{p=1}^{\infty}g_{s}^{2p}\,\frac{B_{2p}}{2p\,(2p-2)!}
\frac{d^{2p}}{dt^{2p}}\,\log R(t, g_{s}), 
\ee
which is equivalent to 
\be
\la{3.36}
F(t+g_{s})-2\,F(t)+F(t-g_{s}) = \log R(t).
\ee
Notice that (\ref{3.35}) and (\ref{3.36}) are weaker than (\ref{3.34}) since they determine 
essentially $F''$ and miss a linear contribution in $t$. In our context this is bad since the coefficients
of this linear part may and do depend on $\rho$ forcing us to use (\ref{3.34}) with the non-trivial
$\rho$ dependent
contribution from $h_{0}$. \footnote{In the later study of 
non-perturbative corrections this will not be an issue
since an algebraic contribution to $F$ cannot mix with a non-perturbative one.}

In the gapped phase we are considering, the function $R$ has a regular expansion in powers of $g_{s}$ plus
non perturbative contributions. The perturbative part \footnote{For ease of notation we do not emphasize
that we are dealing with the perturbative part of $R$ and use $R\equiv R^{\rm pert}$.} is thus
\be
\la{3.37}
R(z, g_{s}) = \sum_{\ell=0}^{\infty}R_{\ell}(z) \,g_{s}^{\ell}.
\ee
The coefficients $R_{\ell}$ may be expanded and computed at first order in $\rho$
\be
R_{\ell}(z) = R_{\ell,0}(z)+R_{\ell,1}(z)\,\rho+ \mc O(\rho^{2}).
\ee
The $\rho$ independent part $R_{\ell,0}(z)$ is well-known, see again \cite{Marino:2008ya},
 and is non zero only for even $\ell$. The first terms read
\begin{align}
R_{2,0}(z) = \frac{1}{8}\frac{z}{(1-z)^{2}},\quad R_{4,0}(z) = \frac{9z(z+3)}{128(1-z)^{5}}
\quad R_{6,0}(z) = \frac{9 z \left(17 z^2+152 z+125\right)}{1024 (1-z)^8},
\end{align}
and so on.
For the $k=2$ and $k=3$ Wilson loops we get respectively 
\begin{align}
& R_{1,1}^{k=2}(z) = z (3 z-2), && R_{3,1}^{k=2}(z) = \frac{z^2 (3 z-5)}{8 (z-1)^3},\notag \\
& R_{5,1}^{k=2}(z) = -\frac{9 (z-3) z^2 (3 z+7)}{128 (z-1)^6}, && 
R_{7,1}^{k=2}(z)=\frac{27 z^2 \left(17 z^3+39 z^2-465 z-375\right)}{1024 (z-1)^9}, \dots
\end{align}
and
\begin{align}
& R_{1,1}^{k=3}(z) = -\frac{3}{2} z \left(10 z^2-12 z+3\right), \quad  R_{3,1}^{k=3}(z) =
-\frac{3 z \left(30 z^3-82 z^2+75 z-25\right)}{16 (z-1)^3} ,\notag \\
& R_{5,1}^{k=3}(z) = \frac{3 z \left(50 z^4+212 z^3-963 z^2+766 z-245\right)}{256 (z-1)^6},\notag \\
& R_{7,1}^{k=3}(z) = -\frac{27 z \left(150 z^5+826 z^4-3483 z^3-1005 z^2+2105 z-945\right)}{2048 (z-1)^9}, \dots
\end{align}
The part of $\log h_{0}$ that is linear in $\rho$   is \footnote{We normalize 
$h_{0}$ by its value at $g_{s}=0$.}
\be
\left. \partial_{\rho}\log h_{0}\right|_{\rho=0} = 
\frac{1}{2}\bigg[\frac{I_{k}(1/g_{s})}{I_{0}(1/g_{s})}-1\bigg],
\ee
and may be easily expanded at small $g_{s}$. Replacing in the general formula (\ref{3.34})
the explicit coefficients $R_{\ell}(z)$ appearing in (\ref{3.37}),  we fully reproduce the 
genus 5 expansion obtained with the resolvent method. \footnote{We recall the small $g_{s}$ expansion
reconstruct the large $N$ expansion. Besides, $t=1/g$ in (\ref{3.34}).}
The advantage of this procedure is that the 
functional equation (\ref{3.33}) -- or similar for higher $k$ -- may be easily and systematically 
expanded in $g_{s}$ at any 
desired order by minor effort thus generating the full $1/N$ genus expansion.

\subsection{Analytical bootstrap from finite $N$ data}
\la{sec:boot}

To conclude our discussion of the perturbative genus expansion in the gapped phase, we present
an analytical bootstrap method that quite efficiently exploits finite $N$ data to general long $1/N$
series for the winding Wilson loops $W_{k}$. This is a sort of educated trick whose great advantage is 
computational simplicity and flexibility. Indeed, it can be applied to more general observables like 
generic Wilson loops in higher ``small'' representations. 

The long genus expansions that we have provided in, {\em e.g.} (\ref{3.17}) and (\ref{3.18}) , 
allow to identify 
strong regularities in how  the coefficients $W_{k,\ell}(g)$ in (\ref{2.12})
depend on $g$.
These regularities were first postulated in \cite{Bessis:1980ss} for the hermitian matrix models
and may be observed in the GWW model as well.
As a warm-up, consider for instance the expansion of $W_{1}^{\rm pert}$
in (\ref{3.15}). It 
apparently takes the form 
\be
\la{3.43}
W_{1}^{\rm pert} = 1-\frac{1}{2\,g}+\frac{1}{N^{2}}\,\frac{1}{8\,g\,(1-g)}+\sum_{p=2}^{\infty}
\frac{1}{N^{2p}}\frac{1}{(g-1)^{3p-2}}\,
\sum_{n=0}^{p-2}c_{p,n}\,g^{n}.
\ee
Since we are in the gapped phase, it makes sense to expand at large $g$, and this gives
the re-organized expansion 
\begin{align}
\la{3.44}
W_{1}^{\rm pert} &= 1-\frac{1}{2 g}-\frac{1}{8  N^2}\frac{1}{g^{2}}-\frac{1}{8 
   N^2}\frac{1}{g^{3}}+\bigg(\frac{c_{2,0}}{N^4}-\frac{1}{8 N^2}\bigg)\,\frac{1}{g^4}
   +\bigg(\frac{4
   c_{2,0}}{N^4}-\frac{1}{8 N^2}\bigg)\,\frac{1}{g^{5}}\notag \\
   & +\bigg(\frac{c_{3,1}}{N^6}+\frac{10
   c_{2,0}}{N^4}-\frac{1}{8 N^2}\bigg)\,\frac{1}{g^6}+\bigg(
   \frac{c_{3,0}}{N^6}+\frac{7
   c_{3,1}}{N^6}+\frac{20 c_{2,0}}{N^4}-\frac{1}{8
   N^2}\bigg)\,\frac{1}{g^{7}}\notag \\
   & +\bigg(\frac{c_{4,2}}{N^8}+\frac{7 c_{3,0}}{N^6}+\frac{28
   c_{3,1}}{N^6}+\frac{35 c_{2,0}}{N^4}-\frac{1}{8
   N^2}\bigg)\,\frac{1}{g^8}+\dots.
\end{align}
The simple observation is that each power of $1/g$ receives contributions from a finite number of terms 
coming from the $1/N$
expansion. This implies that it is possible to match them  with the large $g$ expansion of the 
exact Wilson loop at finite $N$. This is easily obtained by replacing in (\ref{2.1}) or (\ref{2.2})
the modified Bessel function by its asymptotic expansion at large argument
\be
I_{\nu}(x) \sim \frac{e^{x}}{\sqrt{2\,\pi\,x}}\sum_{k=0}^{\infty}
\frac{(\frac{1}{2}+\nu)_{k}(\frac{1}{2}-\nu)_{k}}{k!}\,(2\,x)^{-k}.
\ee
After this replacement, all exponential factors cancel and we get an (asymptotic) 
series in $1/g$ whose coefficients 
depend on $N$. Some explicit examples are 
\begin{align}
\la{3.46}
W_{1}^{(N=2)} &= 1-\frac{1}{2 \,g}-\frac{1}{32 \,g^2}-\frac{1}{32 \,g^3}-\frac{73}{2048 \,g^4}-\frac{25}{512
   \,g^5}-\frac{5153}{65536 \,g^6}-\frac{149}{1024 \,g^7}-\frac{2550997}{8388608
   \,g^8}+\dots,\notag \\
W_{1}^{(N=3)} &= 1-\frac{1}{2 \,g}-\frac{1}{72 \,g^2}-\frac{1}{72 \,g^3}-\frac{17}{1152 \,g^4}-\frac{5}{288
   \,g^5}-\frac{1897}{82944 \,g^6}-\frac{29}{864 \,g^7}-\frac{1299533}{23887872
   \,g^8}+\dots,\notag \\
W_{1}^{(N=4)} &= 1-\frac{1}{2 \,g}-\frac{1}{128 \,g^2}-\frac{1}{128 \,g^3}-\frac{265}{32768
   \,g^4}-\frac{73}{8192 \,g^5}-\frac{44513}{4194304 \,g^6}-\frac{899}{65536
   \,g^7}+\dots,
\end{align}
and so on. Of course, the calculation leading to (\ref{3.46}) is easy and fully straightforward using any
computer algebra system.
Any of the above expansions may be compared with (\ref{3.44}) and this gives 
immediately $c_{2,0}=-9/128$. In a similar way, 
the coefficient of $1/g^{6}$  gives $c_{3,1}=-225/1024$
and so on. It goes without saying that these values  agree with the previous expansions.
 If needed, one may add the information from different $N$ to obtain further constraints. 
To see a less trivial  example, consider the expansion of $W_{3}^{\rm pert}$. 
From inspection of   (\ref{3.17}), our Ansatz is now
\be
\la{3.47}
W_{3}^{\rm pert} = \frac{(g-1)^{2}(2g-5)}{2g^{3}}+\sum_{p=1}^{\infty}\frac{1}{N^{2p}}\frac{1}
{(g-1)^{3p-2}\,g^{3}}\sum_{n=0}^{p+1}d_{p,n}g^{n}.
\ee
Expansion of (\ref{3.47}) at large $g$ reads
\begin{align}
\la{3.48}
W_{3}^{\rm pert} &=
1-\frac{9}{2
   g}+\bigg(\frac{d_{1,2}}{N^2}+6\bigg)\frac{1}{g^{2}}+
   \bigg(\frac{d_{1,1}+d_{1,2}}{N^2
   }-\frac{5}{2}\bigg)\frac{1}{g^3}
   +\bigg(\frac{d_{2,3}}{N^4}+\frac{d_{1,0}+d_{1,1}+d_{1,2}}{N^2}\bigg)\frac{1}{g^4}\notag \\
   &+\bigg(\frac{d_{2,2}+4
   d_{2,3}}{N^4}+\frac{d_{1,0}+d_{1,1}+d_{1,2}}{N^2}\bigg)\frac{1}{g^5}+
   \notag \\
   &+\bigg(\frac{d_{3,4}}{N^6}+\frac{d_{2,1}+4 d_{2,2}+10
   d_{2,3}}{N^4}+\frac{d_{1,0}+d_{1,1}+d_{1,2}}{N^2}\bigg)\frac{1}{g^6}
   \notag \\
   & +\bigg(\frac{d_{3,3}+7 d_{3,4}}{N^6}+\frac{d_{2,0}+4
   d_{2,1}+10 d_{2,2}+20
   d_{2,3}}{N^4}+\frac{d_{1,0}+d_{1,1}+d_{1,2}}{N^2}\bigg)\frac{1}{g^7}+
   \dots.
\end{align}
Explicit expansion of $W_{3}$ at (small !) $N=2,3$ gives 
\begin{align}
W_{3}^{(N=2)} &= 1-\frac{9}{2 g}+\frac{207}{32 g^2}-\frac{93}{32 g^3}-\frac{297}{2048
   g^4}-\frac{81}{512 g^5}-\frac{12465}{65536 g^6}-\frac{567}{2048
   g^7}+\dots, \notag  \\
W_{3}^{(N=3)} &= 1-\frac{9}{2 g}+\frac{149}{24 g^2}-\frac{193}{72 g^3}-\frac{179}{3456
   g^4}-\frac{47}{864 g^5}-\frac{545}{9216 g^6}-\frac{61}{864
   g^7}+\dots.
\end{align}
Again, 
from any of the two expansions we immediately obtain $d_{1,2}=\frac{15}{8}$ from the $1/g^{2}$ term
after comparing with (\ref{3.48}). Then, 
the $1/g^{3}$ term gives $d_{1,1}=-\frac{7}{2}$. Going further, we see the need for combining
different values of $N$. The $1/g^{4}$  term contains 
$d_{2,3}$ and $d_{1,0}$. Taking $N=2,3$ we have 
\be
\left\{\begin{array}{ll}
\frac{1}{4}\,d_{1,0}+\frac{1}{16}\,d_{2,3}-\frac{535}{2048} = 0,\phantom{\bigg(}\\ 
\frac{1}{9}\,d_{1,0}+\frac{1}{81}\,d_{2,3}-\frac{445}{3456} = 0,
\end{array}\right. \qquad\to
d_{1,0}=\frac{5}{4},\ d_{2,3}=-\frac{105}{128}.
\ee
Repeating this procedure, it is possible to generate very long expansions with modest effort. In the 
case of $W_{3}^{\rm pert}$ some additional terms beside those in (\ref{3.17}) are 
\begin{align}
W_{3}^{\rm pert} &= (\ref{3.17}) -\frac{1}{N^{12}}\,\frac{27}{4194304 (g-1)^{16} g^2
}\bigg(
   165540375 g^6+188090000 g^5-936177700 g^4\notag \\
   & +666889840 g^3+2236950835
   g^2+776154848 g+49495570\bigg)\notag \\
   & -\frac{1}{N^{14}}\frac{9}{33554432 (g-1)^{19} g^2}\bigg(119401907625 g^7+336551851125 g^6-742212683100
   g^5\notag \\
   & -227173796100 g^4+2921653890865 g^3+2287870500485 g^2+447179696898
   g+18974970570\bigg)\notag \\
   & 
   -\frac{1}{N^{18}}\,\frac{9}{2147483648
   (g-1)^{22} g^2}\bigg(
324414983017125 g^8+1515451927799250 g^7\notag \\
& -1419779351618775
   g^6-3758360112783000 g^5
    +10443865995456525 g^4\notag \\
    & +16736994071740090
   g^3+6773196057478535 g^2+860519993027628 g+26004983823870\bigg)+\dots.
\end{align}
These may be checked using the method of Sec.~(\ref{sec:PS}). Going for instance to $k=4$, 
a suitable Ansatz similar to (\ref{3.43}) and (\ref{3.47}) gives  immediately  the expansion 
\begin{align}
W_{4}^{\rm pert} &= (\ref{3.18})
+\frac{1}{N^{12}}\frac{9}{1048576 (g-1)^{16} g^3} \bigg(915449850 g^6+1442197600 g^5-4771969775 g^4\notag \\
& -1398884592
   g^3+4976942320 g^2+1870743600 g+112464765\bigg)\notag \\
   & +\frac{1}{N^{14}}\frac{9}{8388608 (g-1)^{19} g^3 }\bigg(220650403050 g^7+734152831650 g^6-1063278944625
   g^5\notag \\
   & -1866656554125 g^4+1544831519736 g^3+1776224765440 g^2+360024742575
   g+14800015755\bigg)\notag \\
   & +\frac{1}{N^{16}}\frac{9}{536870912
   (g-1)^{22} g^3 }\bigg(601691029389450 g^8+3150176313406500 g^7\notag \\
   & -1061039224800225
   g^6-10601097622230350 g^5+854473901206525 g^4\notag \\
   & +11474449077990096
   g^3+5365340415315665 g^2+696133003635450 g+20642622067305\bigg)+\dots\ ,
   \end{align}
and so on for higher $k$. As a check, this long expansions respect the exact constraints
(\ref{kuz1}).

The flexibility of the method can be appreciated by considering different observables related to 
winding Wilson loops and possessing a regular large $N$ expansion like, for instance, the
multi-trace expectation values
\be
W_{k_{1}, \dots, k_{p}} = \langle \prod_{n=1}^{p}\text{tr}(U^{k_{n})}\rangle_{c}.
\ee 
Just to give an example, let us consider
\be
\la{3.54}
W_{1,2}=\langle\text{tr}(U)\,\text{tr}(U^{2})\rangle_{c} = 
\langle\text{tr}(U)\,\text{tr}(U^{2})\rangle-\langle\text{tr}(U)\rangle\langle\text{tr}(U^{2})\rangle.
\ee
The first term in the r.h.s. may be expressed as a difference of Wilson loops in definite 
representations labeled as usual by their Young tableau
\be
\langle\text{tr}(U)\,\text{tr}(U^{2})\rangle = {\tiny \yng(3)}-{\tiny\yng(1,1,1)} = 
\langle\text{tr}_{(3)}(U)\rangle-\langle\text{tr}_{(1,1,1)}(U)\rangle.
\ee
The Wilson loop for the representation $\bm{\lambda}$ is given by 
\be
\langle \text{tr}_{\bm\lambda}(U)\rangle = \frac{\det[I_{\lambda_{m}+n-m}(N\,g)]}
{\det[I_{n-m}(N\,g)]}.
\ee
An educated Ansatz for $W_{1,2}^{\rm pert}$ in (\ref{3.54}) is 
\be
\la{3.57}
W_{1,2}^{\rm pert} = -\frac{2\,(g-1)(g-2)}{g^{3}}+\sum_{p=1}^{\infty}
\frac{1}{N^{2p}}\frac{1}{(g-1)^{3p-1}\,g^{3}}\sum_{n=0}^{p}c_{p,n}g^{n}.
\ee
As before, we compare this Ansatz with the $N=3,4,5,6$ large $g$ expansion of the
exact $W_{1,2}(N, g)$ and we quickly find 
\begin{align}
W_{1,2}^{\rm pert} &= 
-\frac{2 (g-2) (g-1)}{g^3}+\frac{1}{N^{2}}\frac{4-5 g}{4 (g-1)^2 g^3}-\frac{1}{N^{4}}
\frac{9 (7 g-3)}{64 (g-1)^5 g^2}\notag \\
& -\frac{1}{N^{6}}\frac{27
   \left(75 g^2+40 g-17\right)}{512 (g-1)^8 g^2}-\frac{1}{N^{8}}
   \frac{9 \left(67375 g^3+122875 g^2+9453
   g-7143\right)}{16384 (g-1)^{11} g^2}\notag \\
   & -\frac{1}{N^{10}}\frac{9 \left(9029475 g^4+30252950 g^3+17238950 g^2-331818
   g-422781\right)}{131072 (g-1)^{14} g^2}\notag \\
   &-\frac{1}{N^{12}}\frac{27}{2097152 (g-1)^{17} g^2} \bigg(1260653625 g^5+6386462775 g^4+7345350650
   g^3\notag \\
   &+1917278150 g^2-100733611 g-25311493\bigg)\notag \\
   & -\frac{1}{N^{14}}\frac{9}{16777216 (g-1)^{20} g^2}\bigg(
   1138298186025 g^6+7899660650700 g^5\notag \\
   & +14325534191475 g^4
   +8254338677000 g^3+1165668260375 g^2\notag \\
   & -79633801668
   g-9639354243\bigg)+\dots.
\end{align}
Unfortunately, there is no such a trick available for the study of the instanton corrections
in any of the two phases. So, the moral of this section is that the genus expansion in the gapped
phase is somewhat an easy piece. Nevertheless, we have to keep in mind that the starting point 
is an educated guess for the $g$ dependence of the coefficients in (\ref{2.12}), see
(\ref{3.43},\ref{3.47}, \ref{3.57}).  Such expressions should be proved 
rigorously or, at least, formulated from a large set of data points like those provided by the analytical 
algorithms discussed in Sec.~(\ref{sec:resolvent}) or Sec.~(\ref{sec:PS}).

\section{Instanton corrections in the gapless phase}
\la{sec:instanton}

In this section, we discuss how the method proposed in Sec.~(\ref{sec:PS}) may be 
applied to compute the instanton contribution $W_{k}^{\rm inst}$ in (\ref{2.13}).
The basic idea has been put forward in \cite{Marino:2008ya}. We study the gapless phase
as an explicit example. This is particularly interesting because the perturbative corrections
are trivial. \footnote{This implies that we shall be able to test numerically the derived expansions 
in a rather clean way.}

\subsection{Instanton corrections from the Perival-Shevitz recursion: Free energy}

As a warm-up and to recall the basic ideas, let us begin with the free energy without 
sources coupled to $U^{k}$ terms.
In the gapless phase, the Wilson loops $W_{k>1}$ have zero perturbative part and a non trivial
sum of instanton contributions. Let us consider the continuum limit of the Periwal-Shevitz recursion 
for $k=1$, see (\ref{3.33}), 
and write
\be
\la{4.1}
R(z, g_{s}) = 1+\sum_{\ell=1}^{\infty}\mc{C}^{\ell}R^{(\ell)}(z, g_{s}),
\ee
where the $\ell$-instanton contribution $R^{(\ell)}(z, g_{s})$  is written as a trans-series
Ansatz
\be
\la{4.2}
R^{(\ell)}(z, g_{s}) = e^{-\ell\,A(z)/g_{s}}\,R^{(\ell)}_{0}(z)\bigg(1+\sum_{n=1}^{\infty}
g_{s}^{n}\,R^{(\ell)}_{n}(z)\bigg).
\ee
The leading order provides (twice) the instanton action 
\be
A(z) = 2\,z\,\cosh^{-1}(z)-2\,\sqrt{z^{2}-1},\qquad z>1.
\ee
For the leading contribution $\ell=1$, we readily compute the expansion functions in (\ref{4.2})
\begin{align}
R^{(1)}_{0}(z) &= \frac{1}{\sqrt{z^{2}-1}}, \qquad
R^{(1)}_{1}(z) =-\frac{2 z^2+3}{12 \left(z^2-1\right)^{3/2}}, \notag \\
R^{(1)}_{2}(z) &=\frac{4 z^4+156 z^2+45}{288 (z-1)^3 (z+1)^3}, \qquad 
R^{(1)}_{3}(z) =\frac{248 z^6-31716 z^4-73602 z^2-8505}{51840 \left(z^2-1\right)^{9/2}}, \notag \\
R^{(1)}_{4}(z) &=-\frac{2224 z^8-1476192 z^6-12418488 z^4-10071864 z^2-616005}{2488320 (z-1)^6
   (z+1)^6}.
\end{align}
Plugging these in (\ref{4.2}) and using (\ref{3.36}) we obtain the free energy. Finally, using the
relation (\ref{2.3}) with the constant $\mc C$ fixed in \cite{Okuyama:2017pil}, \footnote{
\la{foot1}
This is clearly an important issue. In general, the trans-series Ansatz (\ref{4.1}) contains the 
free parameter $\mc C$,
cf. (\ref{4.1})
This may be fixed by a careful comparison with the double scaling limit of the matrix model, 
see for instance the nice computation in App. C of \cite{Okuyama:2017pil}. We shall adopt the choice
of $\mc C$ in that reference for all $W_{k}$. 
We have checked numerically at high precision its correctness in all considered cases. Notice that
this is rigorous for the cases in (\ref{kuz1}).
} 
we get the leading instanton contribution to $W_{1}^{\rm inst}$ that may be written as
\begin{align}
\la{4.5}
W_{1}^{\rm inst} &= \frac{e^{-2\,N\,S(g)}}{4\,\pi\,N^{2}}\,\bigg[
\frac{g}{g^2-1}+\frac{1}{N}\frac{g \left(3 g^2+14\right)}{12 \left(1-g^2\right)^{5/2}
   }-\frac{1}{N^{2}}\frac{g \left(81 g^4+804 g^2+340\right)}{288 \left(g^2-1\right)^4
   }\notag \\
   & +\frac{1}{N^{3}}\frac{g \left(23085 g^6+406782 g^4+504756 g^2+60952\right)}{51840
   \left(1-g^2\right)^{11/2} }\notag \\
   & +\frac{1}{N^{4}}\frac{g \left(2278125 g^8+63318024 g^6+155794104
   g^4+63098400 g^2+2923472\right)}{2488320 \left(g^2-1\right)^7}+\dots
\bigg],
\end{align}
where the 1-instanton action $S(g)$ has been defined  in (\ref{2.14}).

\subsection{Instanton corrections to winding Wilson loops}
\la{sec:inst-wind}

For a winding loop the procedure is completely similar. We start with the $k$-th Periwal-Shevitz
recursion and work out the $R_{n}$ coefficients at first order in the source $\rho$.
For example, in the case of $W_{2}^{\rm inst}$, we obtain 
\be
R^{(1), k=2}_{0}(z; \rho) = \frac{1-2\,z\,\sqrt{z^{2}-1}\,\rho+\mc O(\rho^{2})}{\sqrt{z^{2}-1}},
\ee
and, writing $R^{(1)}_{n}(z; \rho) = R^{(1)}_{n}(z; 0) + H^{(1)}_{n}(z)\,\rho+\mc O(\rho)^{2}$,
the next coefficient functions are 
\begin{align}
\la{4.7}
H^{(1), k=2}_{1}(z) &=z \left(\frac{1}{1-z^2}+2\right), \qquad
H^{(1),k=2}_{2}(z) =-\frac{z \left(z^4-3 z^2-3\right)}{3 \left(z^2-1\right)^{5/2}}, \notag \\
H^{(1),k=2}_{3}(z) &=\frac{z \left(8 z^6-36 z^4-1182 z^2-225\right)}{288 \left(z^2-1\right)^4}, \notag \\
H^{(1),k=2}_{4}(z) &=\frac{z \left(248 z^8-1488 z^6+232182 z^4+311418 z^2+25515\right)}{25920
   \left(z^2-1\right)^{11/2}}.
\end{align}
These values give an expression of $W_{2}^{\rm inst}$ that is in full agreement with (\ref{2.7})
taking into account the result (\ref{4.5}) for $W_{1}^{\rm inst}$. This is a check and had to be true
because the relation (\ref{2.7}) is exact at finite $N,g$, so it remains valid in any special limit and 
separately for the perturbative and non-perturbative corrections.

A subtle point in the procedure is that the functions $H_{n}^{(1)}(z)$ are obtained
integrating a first order differential equation. The integration constant can be absorbed in the
overall constant $\mc C$, see (\ref{4.1}). As we mentioned before, in the case of the free energy
we exploited the calculations of \cite{Okuyama:2017pil} to fix it at $\rho=0$. Here, 
$\mc C = \mc C(\rho)$ and there may be 
a contribution to $\mc C$ linear in $\rho$. In the case of even winding loops $W_{2k}$ this is not an issue
since integration constants in (\ref{4.7}) are forbidden by parity under $z\to -z$. This explains why
the agreement with (\ref{2.7}) is not accidental. For odd winding loops $W_{2k+1}$ this is 
potentially a serious 
obstacle. In App. ~(\ref{App:W3}) we illustrate a numerical procedure to fix the 
$\rho$ dependent part of the Stokes constant $\mc C$. Actually, we shall return on this issue in 
Sec.~(\ref{sec:general})) showing that the integration constant problem may be 
solved by constructing the expansion for generic even $k$ and simply continuing the 
resulting expressions to odd $k$.

\begin{figure}[t]  
  \centering
\includegraphics[scale=0.35]{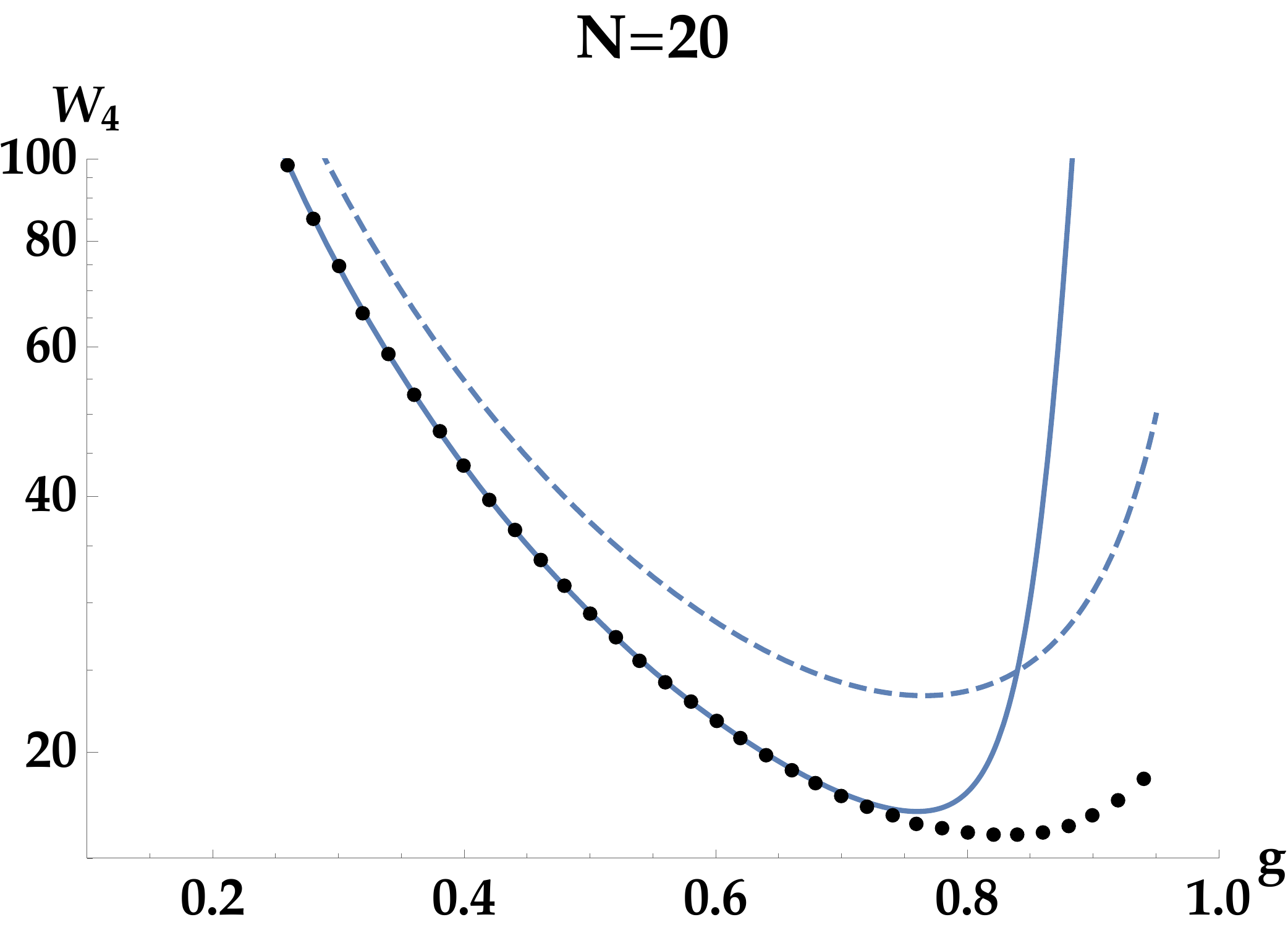}
\includegraphics[scale=0.35]{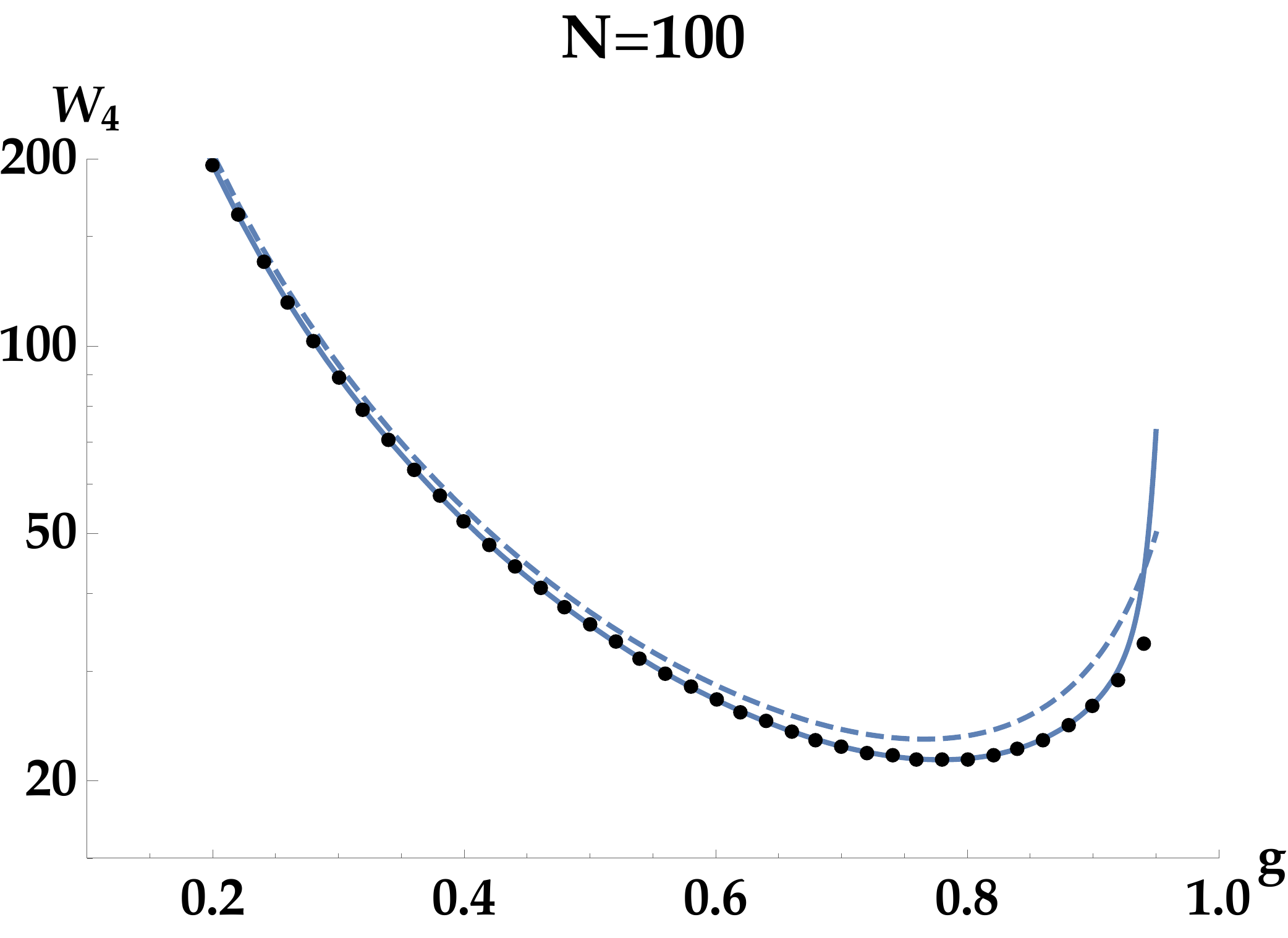}
   \caption{
   Numerical evaluation of the Wilson loop $\widehat W_{4}$ (the hat means that we stripped 
   off the 
   exponential prefactor as explained in the text). Left and right panels are for $N=20,100$. Black
   dots show the exact high precision numerical evaluation of the exact loop. The solid blue line
   is the full expansion (\ref{4.8}), while the dashed line is its lowest leading order at large $N$.
   The vertical scale is logarithmic to improve data visualization.
    \label{fig1}  }
\end{figure}

The next non-trivial even loop is $W_{4}$. In this case  we have a 94-term Periwal-Shevitz recursion,
see App.~(\ref{app:PS4}). Applying the above procedure,
the final expansion of the Wilson loop reads
\begin{align}
\la{4.8}
W_{4}^{\rm inst} &= \frac{e^{-2\,N\,S(g)}}{4\,\pi\,N^{2}}\,\bigg[
\frac{8-4\,g^{2}}{(1-g^{2})\,g^{2}}
-\frac{1}{N}\frac{69 g^4-152 g^2+100}{3 g^2
   \left(1-g^2\right)^{5/2}}-\frac{1}{N^{2}}\frac{2529 g^6-7422 g^4+7516 g^2-3848}{72 g^2
   \left(g^2-1\right)^4}\notag \\
   & +\frac{1}{N^{3}}\frac{104085 g^8+505332 g^6-1450368 g^4+576400
   g^2-731024}{12960  g^2 (1-g^2)^{11/2}}\notag \\
  & -\frac{1}{N^{4}}\frac{-6827085
   g^{10}-90933678 g^8+98056440 g^6+138076944 g^4+113887984 g^2+35151520}{622080 g^2
   (g^2-1)^7 }+\dots
\bigg].
\end{align}
One can check that this expansion is in agreement with the leading instanton order of the exact
relation (\ref{kuz1}).
To appreciate the accuracy of (\ref{4.8}), we compare in Fig.~(\ref{fig1}) the numerical 
evaluation of $W_{4}$ with its expansion (\ref{4.8}). We plot $\widehat W_{4}$ which strips off the exponential prefactor $W_{4} = \frac{e^{-2\,N\,S(g)}}{4\,\pi\,N^{2}}\,\widehat{W}_{4}$. In the figure,
black dots show the high precision numerical evaluation of the Wilson loop, the solid blue line is the 
whole expression in (\ref{4.8}), while the dashed blue line is the leading order, {\em i.e.}
the $\frac{8-4\,g^{2}}{(1-g^{2})g^{2}}$ term alone.
Already at $N=20$ (left panel), the expansion (\ref{4.8}) is quite accurate up to $g\simeq 0.7$ where the expansion
starts breaking down. The $1/N$ corrections are definitely important to achieve this. For higher $N=100$
(right panel) the accuracy of (\ref{4.8}) extends to a wider range. The relative role of $1/N$ corrections
is obviously less visible, but clearly relevant.

\subsection{A general formula for the next-to-leading order expansion of general $W_{k}^{\rm inst}$}
\la{sec:general}

The fact that (\ref{4.1}) starts with $R=1+\dots$ has an important consequence. At the leading
instanton level, the only terms in the $k$-th Periwal-Shevitz recursion that play a role are 
those linear in $f$, {\em i.e.}
the pair $f_{n+k}+f_{n-k}$, see (\ref{3.30}), (\ref{3.31}) and (\ref{B.1}). This means that we
can work out the procedure illustrated in Sec.~(\ref{sec:inst-wind}) parametrically in $k$. 
This is a major simplification given
the growing  complexity of the recursion with increasing $k$
At leading order in the $g_{s}$ expansion, we obtain 
\be
\la{4.9}
R_{0}^{(1), k}(z; \rho) = \frac{1}{\sqrt{z^{2}-1}}\,\bigg[
1+(-1)^{k}\sinh(k\,\cosh^{-1}(z)) \,\rho+ \mc O(\rho^{2})\bigg].
\ee
From this expression, a straightforward calculation gives the generalization of the leading term in 
(\ref{4.5},\ref{A.3},\ref{4.8}) to all $k$. It reads
\begin{align}
\la{4.10}
W_{k}^{\rm inst} &= \frac{e^{-2\,N\,S(g)}}{4\,\pi\,N^{2}}\,\bigg[(-1)^{k}\frac{g}{1-g^{2}}
\,\mathscr U_{k-1}\left(\frac{1}{g}\right) + \mc O(N^{-1})\bigg],
\end{align}
where $\mathscr U_{n}(x)$ are Chebyshev polynomials of the second kind. \footnote{
Useful relations for the Chebyshev polynomials that can be used for $0<g<1$ are  
\be\notag 
\sinh[k\,\text{sech}^{-1}(g)] = g^{-1}(1-g^{2})^{1/2}\mathscr{U}_{k-1}(g^{-1}),\quad
\cosh[k\,\text{sech}^{-1}(g)] = \mathscr{T}_{k}(g^{-1}).
\ee
}
The compact relation (\ref{4.10}) can be checked against  (\ref{2.6}). Indeed, 
at the leading instanton level we can simply replace 
\be
w_{1} = \frac{g}{2}+\frac{e^{-2\,N\,S(g)}}{4\,\pi\,N^{2}}\,\frac{g}{g^{2}-1}+\cdots, \quad
w_{k\ge 2} = \frac{e^{-2\,N\,S(g)}}{4\,\pi\,N^{2}}\,\widehat w_{k}+\cdots,
\ee
and we get 
\begin{align}
\sum_{n=2}^{\infty}\widehat{w}_{n}\,x^{n} &= \frac{g\,x^{2}\,(2+g\,x)}{(1-g^{2})(g+2x+gx^{2})},
\end{align}
which is indeed equivalent to (\ref{4.10}) using the generating function of Chebyshev
polynomials. The reason why we can use (\ref{2.6}) is once again that subleading corrections
to large $N$ factorization  do not affect the leading term in (\ref{4.10}).

Further corrections 
in $1/N$ to (\ref{4.10}) can also be worked out starting the perturbative machinery from the initial seed (\ref{4.9}). Remarkably, this may be done for a generic even  $k$ and continued to odd $k$.  
After some tedious manipulation, the general expression for the first two corrections to (\ref{4.10})
is in full generality \footnote{The drastic simplification of the Periwal-Shevitz recursion at leading
instanton order also means that it can be solved in closed form in terms of the Debye expansion
of Bessel functions, as illustrated in App. C of  \cite{Okuyama:2017pil} for the $k=1$ recursion
relevant for the GWW free energy. We do not pursue this remark because the observable Wilson 
loops still require the inversion of (\ref{3.36}). This gives terms like those in (\ref{4.13}) but not in 
explicit closed form.}
\begin{align}
\la{4.13}
W_{k}^{\rm inst} &= (-1)^{k}\,\frac{e^{-2\,N\,S(g)}}{4\,\pi\,N^{2}}\,\bigg\{\frac{g}{1-g^{2}}
\,\mathscr U_{k-1}(g^{-1})\notag \\
& +\frac{1}{N}\,\frac{g}{12\,(1-g^{2})^{5/2}}\bigg[
18\,g\,k\,\mathscr T_{k}(g^{-1})+[3\,(2 k^{2}-3)\,g^{2}-6k^{2}-26]\,\mathscr U_{k-1}(g^{-1})
\bigg]\notag\\
&+\frac{1}{N^{2}}\frac{g}{288\,(1-g^{2})^{4}}\bigg[
24\,g\,k\,[(11k^{2}-53)\,g^{2}-11k^{2}-52]\,\mathscr T_{k}(g^{-1})+\notag \\
& +[9\,(4k^{4}-28k^{2}+33)\,g^{4}-12\,(6k^{4}+50k^{2}-207)\,g^{2}+36k^{4}+852k^{2}+964]\,
\,\mathscr U_{k-1}(g^{-1})
\bigg]\notag \\
& +\mc O(N^{-3})\bigg\},
\end{align}
where $\mathscr T_{k}(x)$ are Chebyshev polynomials of the first kind. 
Explicit evaluation for $k=1,2,4$ reproduces the previous results. 
For $k=3$ we find from (\ref{4.13})
\begin{align}
\la{4.14}
W_{3}^{\rm inst} &=  \frac{e^{-2\,N\,S(g)}}{4\,\pi\,N^{2}}\,\bigg[
\frac{g^{2}-4}{g(1-g^{2})}
+\frac{1}{N}\frac{45 g^4-98 g^2+104}{12 g \left(1-g^2\right)^{5/2}
   }+\frac{1}{N^{2}}\frac{945 g^6-2592 g^4+676 g^2-2704}{288 g (g^2-1)^4}+\dots
\bigg].
\end{align}
The $1/N$ correction automatically agrees with the result derived in App.~(\ref{App:W3})
where we fixed the $\rho$ part of the Stokes factor $\mc C$ by a numerical calculation. Here, it is 
given automatically from the (trivial) continuation of (\ref{4.13})
from even $k$ to odd $k$. Once again, the expansion (\ref{4.14}) is in full agreement with 
(\ref{kuz1}) at leading instanton order.
Similarly, for $W_{5}$ 
we obtain from (\ref{4.13}) the expansion 
\begin{align}
W_{5}^{\rm inst} &=  \frac{e^{-2\,N\,S(g)}}{4\,\pi\,N^{2}}\,\bigg[
-\frac{g^4-12 g^2+16}{g^3 \left(1-g^2\right)}+\frac{1}{N}\frac{-141 g^6+1418 g^4-2568
   g^2+1376}{12 g^3 \left(1-g^2\right)^{5/2} }\notag \\
   & +\frac{1}{N^{2}}\frac{-16497 g^8+122280
   g^6-269908 g^4+246384 g^2-88384}{288 g^3 (g^2-1)^4
   }
+\dots
\bigg],
\end{align}
that we checked numerically agains the exact value of $W_{5}$ with the expected numerical 
accuracy ( the difference goes to zero as $N^{-3}$ for a generic fixed $g$).

\section{Conclusions}

In this paper, we reconsidered the large $N$ expansion of the Gross-Witten-Wadia
 matrix model and, in particular, of the winding Wilson loops 
\mbox{$W_{k}=\frac{1}{N}\langle\text{tr}U^{k}\rangle$}. 
The motivation of our analysis was to test recent numerical results about these expansions.
 We have presented various analytical algorithms that are able to accurately compute  
both the perturbative and instanton corrections in the gapless and gapped phases of the GWW model.
All the derived expansions are in full agreement with the Schwinger-Dyson prediction when
available. 

These tools also apply to the treatment of  Wilson loops in small representation 
where the number of boxes of the associated Young tableau is small compared to $N$.
In principle, they may be useful to discuss the more interesting cases of large representations 
where the number of boxes grows $\sim N$, such those relevant for the study of $k$-strings, 
see for instance \cite{Karczmarek:2010ec}, or giant Wilson loops \cite{Grignani:2009ua}. 
This seems particularly feasible for the instanton corrections. Indeed,  the complexity of the Periwal-Shevitz  recursion increases with $k$, but most of its terms simply 
do not enter such calculation, as we explained in the text. For instance, one of our main
results is the accurate expansion (\ref{4.13}) that is fully parametrical in $k$ and may be the starting 
point for such investigations.
Finally, it could be interesting to study the double scaling limit
of the winding Wilson loops starting from their computation in terms of the modified 
Periwal-Shevitz recursion.

\section*{Acknowledgments}
We thank K. Okuyama for very interesting and clarifying discussions.

\appendix

\section{A technical aside: Fixing the Stokes constant in instanton corrections to odd winding loops}
\la{App:W3}

\begin{figure}[t]  
  \centering
\includegraphics[scale=0.5]{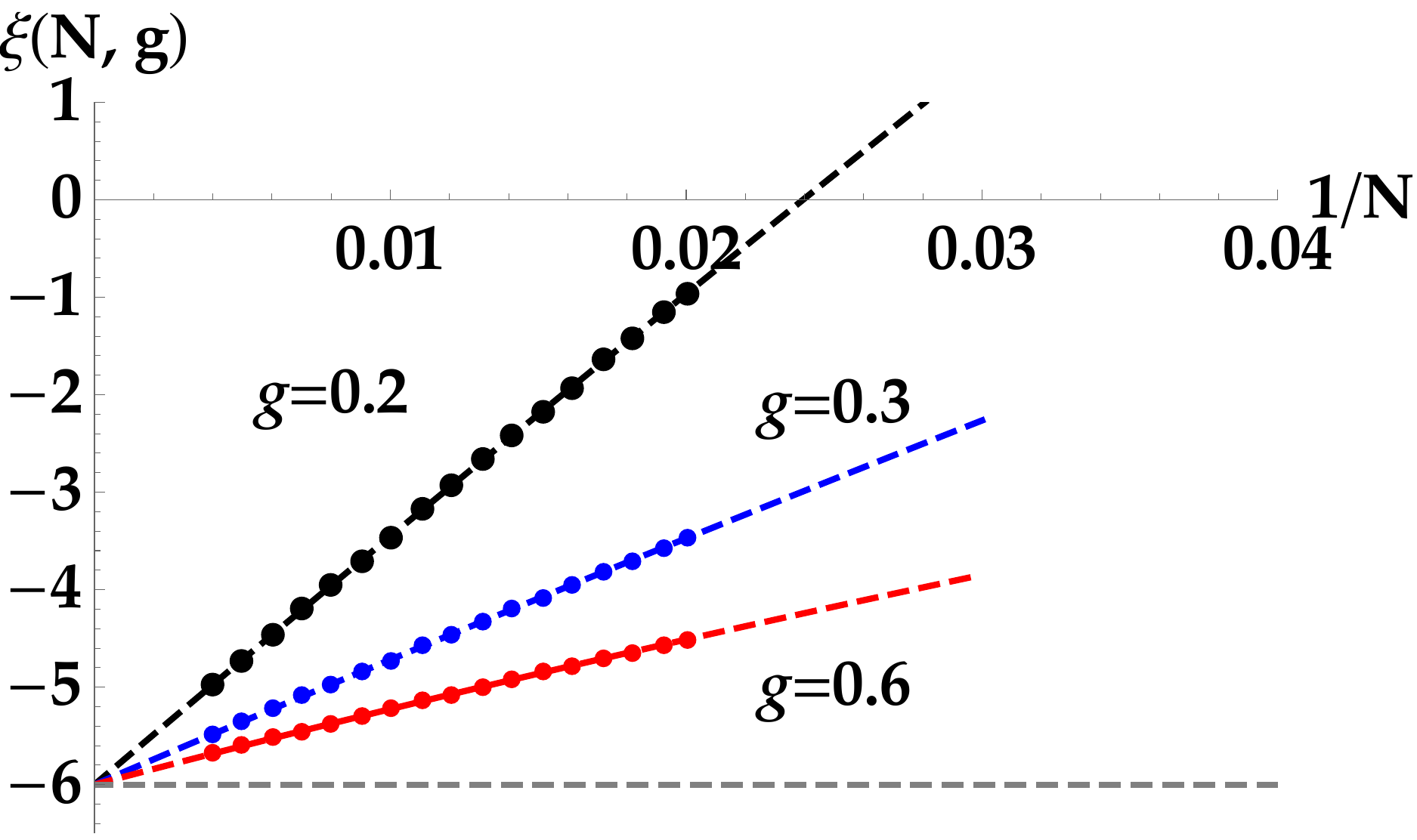}
   \caption{
   Best estimate $\xi(N,g)$ from the exact $W_{3}(N,g)$ at high precision 
   compared with (\ref{A.3}). The three data sets are obtained with 
   $g=0.2\ \text{(black)}, 0.3\ \text{(blue)}, 
0.6\ \text{(red)}$. Extrapolation to $N\to \infty$ is fully consistent with $\xi=-6$ 
(horizontal dashed line)
independently on $g$.
    \label{fig2}  }
\end{figure}
As we commented in  Sec.~(\ref{sec:inst-wind}) the calculation of instanton corrections to odd 
winding loops by the Periwal-Shevitz recursion is plagued by a technical problem. 
Repeating the first steps illustrated in that section in the simplest case of $W_{3}$ we 
first obtain  
\be
R^{(1), k=3}_{0}(z; \rho) = \frac{1+(4z^{2}-1)\,\sqrt{z^{2}-1}\,\rho+\mc O(\rho^{2})}{\sqrt{z^{2}-1}},
\ee
and the first correction at small $g_{s}$ expansion reads
\be
\la{A.2}
H^{(1), k=3}_{1}(z) =-3 \left(4 z^2+\frac{1}{2-2 z^2}-4\right)+\xi,
\ee
where $\xi$ is a numerical constant. The associated expansion of $W_{3}^{\rm inst}$ is 
then obtained from (\ref{3.36}) and reads
\begin{align}
\la{A.3}
W_{3}^{\rm inst} &= \frac{e^{-2\,N\,S(g)}}{4\,\pi\,N^{2}}\,\bigg[
\frac{g^{2}-4}{g(1-g^{2})}+\frac{1}{N}\,
\frac{3 g^4 (4 \xi +39)-2 g^2 (6 \xi +85)+104}{12
   g \left(1-g^2\right)^{5/2}}+\mc O(N^{-2})\bigg].
\end{align}
A non zero constant $\xi$ in (\ref{A.2}) is not forbidden by parity $z\to -z$ and 
inspection of  (\ref{A.3}) does not reveal any 
obvious way to fix it. This means that further non perturbative information
is needed to fix it. A possibility is to compare the exact evaluation of $W_{3}(N,g)$ with the 
2-term expansion in (\ref{A.3}) and determine the optimal $\xi(N,g)$ that gives equality
at the considered $(N,g)$. The correct $\xi$
is obtained as $N\to \infty$ and must be independent on $0<g<1$ as a consistency check.
We show this procedure in Fig.~(\ref{fig2}). We computed the exact $W_{3}(N,g)$ at high precision 
(400 digits)
for $N=50, \dots, 250$ at the three values $g=0.2\ \text{(black)}, 0.3\ \text{(blue)}, 
0.6\ \text{(red)}$. Extrapolation to $N\to \infty$ is fully consistent with $\xi=-6$ with high precision 
(the relative error is $\sim 10^{-6}$).

As explained in Sec.~(\ref{sec:general}), these difficulties and cumbersome procedure
can be automatically overcome starting from expressions valid for generic even $k$ and continuing to 
odd $k$ thus providing a fully analytic result that can be extended to higher $1/N$ accuracy in 
algorithmic way. Of course, in the specific case of $W_{3}$, the exact result (\ref{kuz1})
is fully reproduced.


\section{Periwal-Shevitz recursion relation for $W_{4}$}
\la{app:PS4}

The $k=4$ case of the recursion (\ref{3.27}) is 
\begin{align}
\la{B.1}
k&=4, \quad g_{s}\,(n+1)\,f_{n} = -2 (1-f_{n}^{2})\,
(
f_{n-2}^3 f_{n-1}^4+f_n^3 f_{n-1}^4+3 f_{n-2} f_n^2 f_{n-1}^4+3 f_{n-2}^2 f_n f_{n-1}^4\notag \\
& +2 f_{n-3}
   f_{n-2}^3 f_{n-1}^3-2 f_{n-3} f_{n-2} f_{n-1}^3+2 f_{n-3} f_{n-2}^2 f_n f_{n-1}^3-2 f_{n-3} f_n
   f_{n-1}^3+4 f_n^3 f_{n+1} f_{n-1}^3\notag \\
   & +6 f_{n-2} f_n^2 f_{n+1} f_{n-1}^3-2 f_{n-2} f_{n+1} f_{n-1}^3+2
   f_{n-2}^2 f_n f_{n+1} f_{n-1}^3-2 f_n f_{n+1} f_{n-1}^3+f_{n-3}^2 f_{n-2}^3 f_{n-1}^2\notag \\
   & -f_{n-2}^3
   f_{n-1}^2-f_{n-4} f_{n-3}^2 f_{n-1}^2+f_{n-4} f_{n-3}^2 f_{n-2}^2 f_{n-1}^2-f_{n-4} f_{n-2}^2
   f_{n-1}^2-3 f_{n-2} f_n^2 f_{n-1}^2\notag \\
   & +6 f_n^3 f_{n+1}^2 f_{n-1}^2+3 f_{n-2} f_n^2 f_{n+1}^2
   f_{n-1}^2-f_{n-2} f_{n+1}^2 f_{n-1}^2-4 f_n f_{n+1}^2 f_{n-1}^2+f_{n-4} f_{n-1}^2\notag \\
   & -f_{n-3}^2 f_{n-2}
   f_{n-1}^2-4 f_{n-2}^2 f_n f_{n-1}^2+2 f_{n-3} f_{n-2}^2 f_n f_{n+1} f_{n-1}^2-2 f_{n-3} f_n f_{n+1}
   f_{n-1}^2-3 f_n^2 f_{n+2} f_{n-1}^2\notag \\
   & +3 f_n^2 f_{n+1}^2 f_{n+2} f_{n-1}^2+2 f_{n-2} f_n f_{n+1}^2
   f_{n+2} f_{n-1}^2-f_{n+1}^2 f_{n+2} f_{n-1}^2-2 f_{n-2} f_n f_{n+2} f_{n-1}^2+f_{n+2} f_{n-1}^2
   \notag \\
   & -2
   f_{n-3} f_{n-2}^3 f_{n-1}+4 f_n^3 f_{n+1}^3 f_{n-1} -2 f_n f_{n+1}^3 f_{n-1}+2 f_n f_{n+1}^3 f_{n+2}^2
   f_{n-1}-2 f_n f_{n+1} f_{n+2}^2 f_{n-1}\notag \\
    & +2 f_{n-3} f_{n-2} f_{n-1}-2 f_{n-3} f_{n-2}^2 f_n f_{n-1}+2
   f_{n-3} f_n f_{n-1}-6 f_{n-2} f_n^2 f_{n+1} f_{n-1}+2 f_{n-2} f_{n+1} f_{n-1}\notag \\
   & -2 f_{n-2}^2 f_n f_{n+1}
   f_{n-1}+6 f_n^2 f_{n+1}^3 f_{n+2} f_{n-1}-2 f_{n+1}^3 f_{n+2} f_{n-1}-6 f_n^2 f_{n+1} f_{n+2}
   f_{n-1}+2 f_{n+1} f_{n+2} f_{n-1}\notag \\
   & -2 f_n f_{n+1}^2 f_{n+3} f_{n-1}+2 f_n f_{n+1}^2 f_{n+2}^2 f_{n+3}
   f_{n-1}-2 f_n f_{n+2}^2 f_{n+3} f_{n-1}+2 f_n f_{n+3} f_{n-1}+f_n^3 f_{n+1}^4\notag \\
   & -f_{n-3}^2
   f_{n-2}^3+f_{n+1}^4 f_{n+2}^3-f_{n+1}^2 f_{n+2}^3+f_{n-4} f_{n-3}^2-f_{n-4} f_{n-3}^2
   f_{n-2}^2+f_{n-4} f_{n-2}^2-3 f_{n-2} f_n^2 f_{n+1}^2\notag \\
   & +f_{n-2} f_{n+1}^2+3 f_n f_{n+1}^4 f_{n+2}^2-4
   f_n f_{n+1}^2 f_{n+2}^2+f_n f_{n+2}^2+f_{n+1}^2 f_{n+2}^3 f_{n+3}^2-f_{n+2}^3 f_{n+3}^2
   \notag \\
   & -f_{n+1}^2
   f_{n+2} f_{n+3}^2+f_{n+2} f_{n+3}^2-f_{n-4}+f_{n-3}^2 f_{n-2}+f_{n-2}^2 f_n-2 f_{n-3} f_{n-2}^2 f_n
   f_{n+1}+2 f_{n-3} f_n f_{n+1}\notag \\
   & +3 f_n^2 f_{n+1}^4 f_{n+2}-3 f_n^2 f_{n+1}^2 f_{n+2}-2 f_{n-2} f_n
   f_{n+1}^2 f_{n+2}+2 f_{n-2} f_n f_{n+2}-2 f_n f_{n+1}^3 f_{n+3}+2 f_{n+1}^3 f_{n+2}^3 f_{n+3}
   \notag \\
   & -2
   f_{n+1} f_{n+2}^3 f_{n+3}+2 f_n f_{n+1}^3 f_{n+2}^2 f_{n+3}-2 f_n f_{n+1} f_{n+2}^2 f_{n+3}+2 f_n
   f_{n+1} f_{n+3}-2 f_{n+1}^3 f_{n+2} f_{n+3}\notag \\
   & +2 f_{n+1} f_{n+2} f_{n+3}+f_{n+1}^2 f_{n+4}-f_{n+1}^2
   f_{n+2}^2 f_{n+4}+f_{n+2}^2 f_{n+4}-f_{n+1}^2 f_{n+3}^2 f_{n+4}+f_{n+1}^2 f_{n+2}^2 f_{n+3}^2
   f_{n+4}\notag \\
   & -f_{n+2}^2 f_{n+3}^2 f_{n+4}+f_{n+3}^2 f_{n+4}-f_{n+4}
).
\end{align}

\bibliography{BT-Biblio}

\providecommand{\href}[2]{#2}\begingroup\raggedright\begin{thebibliography}{10}

\bibitem{brezin1993large}
E.~Br{\'e}zin and S.~R. Wadia, \emph{The large N expansion in quantum field
  theory and statistical physics: from spin systems to 2-dimensional gravity}.
\newblock World scientific, 1993.

\bibitem{Rossi:1996hs}
P.~Rossi, M.~Campostrini and E.~Vicari, \emph{{The Large N expansion of unitary
  matrix models}},
  \href{http://dx.doi.org/10.1016/S0370-1573(98)00003-9}{\emph{Phys. Rept.}
  {\bf 302} (1998) 143--209}, [\href{http://arxiv.org/abs/hep-lat/9609003}{{\tt
  hep-lat/9609003}}].

\bibitem{akemann2011oxford}
G.~Akemann, J.~Baik and P.~Di~Francesco, \emph{The Oxford handbook of random
  matrix theory}.
\newblock Oxford University Press, 2011.

\bibitem{wang2013application}
C.~B. Wang, \emph{Application of integrable systems to phase transitions}.
\newblock Springer, 2013.

\bibitem{Stanley:1968gx}
H.~E. Stanley, \emph{{Spherical model as the limit of infinite spin
  dimensionality}},
  \href{http://dx.doi.org/10.1103/PhysRev.176.718}{\emph{Phys. Rev.} {\bf 176}
  (1968) 718--722}.

\bibitem{Ma:1973zu}
S.-k. Ma, \emph{{Introduction to the renormalization group}},
  \href{http://dx.doi.org/10.1103/RevModPhys.45.589}{\emph{Rev. Mod. Phys.}
  {\bf 45} (1973) 589--614}.

\bibitem{tHooft:1973alw}
G.~'t~Hooft, \emph{{A Planar Diagram Theory for Strong Interactions}},
  \href{http://dx.doi.org/10.1016/0550-3213(74)90154-0}{\emph{Nucl. Phys.} {\bf
  B72} (1974) 461}.

\bibitem{Gross:1980he}
D.~J. Gross and E.~Witten, \emph{{Possible Third Order Phase Transition in the
  Large N Lattice Gauge Theory}},
  \href{http://dx.doi.org/10.1103/PhysRevD.21.446}{\emph{Phys. Rev.} {\bf D21}
  (1980) 446--453}.

\bibitem{Erickson:2000af}
J.~K. Erickson, G.~W. Semenoff and K.~Zarembo, \emph{{Wilson loops in N=4
  supersymmetric Yang-Mills theory}},
  \href{http://dx.doi.org/10.1016/S0550-3213(00)00300-X}{\emph{Nucl. Phys.}
  {\bf B582} (2000) 155--175}, [\href{http://arxiv.org/abs/hep-th/0003055}{{\tt
  hep-th/0003055}}].

\bibitem{Drukker:2000rr}
N.~Drukker and D.~J. Gross, \emph{{An Exact prediction of N=4 SUSYM theory for
  string theory}}, \href{http://dx.doi.org/10.1063/1.1372177}{\emph{J. Math.
  Phys.} {\bf 42} (2001) 2896--2914},
  [\href{http://arxiv.org/abs/hep-th/0010274}{{\tt hep-th/0010274}}].

\bibitem{Pestun:2007rz}
V.~Pestun, \emph{{Localization of gauge theory on a four-sphere and
  supersymmetric Wilson loops}},
  \href{http://dx.doi.org/10.1007/s00220-012-1485-0}{\emph{Commun. Math. Phys.}
  {\bf 313} (2012) 71--129}, [\href{http://arxiv.org/abs/0712.2824}{{\tt
  0712.2824}}].

\bibitem{Drukker:2005kx}
N.~Drukker and B.~Fiol, \emph{{All-genus calculation of Wilson loops using
  D-branes}},
  \href{http://dx.doi.org/10.1088/1126-6708/2005/02/010}{\emph{JHEP} {\bf 02}
  (2005) 010}, [\href{http://arxiv.org/abs/hep-th/0501109}{{\tt
  hep-th/0501109}}].

\bibitem{Gomis:2006sb}
J.~Gomis and F.~Passerini, \emph{{Holographic Wilson Loops}},
  \href{http://dx.doi.org/10.1088/1126-6708/2006/08/074}{\emph{JHEP} {\bf 08}
  (2006) 074}, [\href{http://arxiv.org/abs/hep-th/0604007}{{\tt
  hep-th/0604007}}].

\bibitem{Gomis:2006im}
J.~Gomis and F.~Passerini, \emph{{Wilson Loops as D3-Branes}},
  \href{http://dx.doi.org/10.1088/1126-6708/2007/01/097}{\emph{JHEP} {\bf 01}
  (2007) 097}, [\href{http://arxiv.org/abs/hep-th/0612022}{{\tt
  hep-th/0612022}}].

\bibitem{Hartnoll:2006is}
S.~A. Hartnoll and S.~P. Kumar, \emph{{Higher rank Wilson loops from a matrix
  model}}, \href{http://dx.doi.org/10.1088/1126-6708/2006/08/026}{\emph{JHEP}
  {\bf 08} (2006) 026}, [\href{http://arxiv.org/abs/hep-th/0605027}{{\tt
  hep-th/0605027}}].

\bibitem{Okuyama:2006jc}
K.~Okuyama and G.~W. Semenoff, \emph{{Wilson loops in N=4 SYM and fermion
  droplets}},
  \href{http://dx.doi.org/10.1088/1126-6708/2006/06/057}{\emph{JHEP} {\bf 06}
  (2006) 057}, [\href{http://arxiv.org/abs/hep-th/0604209}{{\tt
  hep-th/0604209}}].

\bibitem{Yamaguchi:2007ps}
S.~Yamaguchi, \emph{{Semi-classical open string corrections and symmetric
  Wilson loops}},
  \href{http://dx.doi.org/10.1088/1126-6708/2007/06/073}{\emph{JHEP} {\bf 06}
  (2007) 073}, [\href{http://arxiv.org/abs/hep-th/0701052}{{\tt
  hep-th/0701052}}].

\bibitem{Drukker:2006zk}
N.~Drukker, S.~Giombi, R.~Ricci and D.~Trancanelli, \emph{{On the D3-brane
  description of some 1/4 BPS Wilson loops}},
  \href{http://dx.doi.org/10.1088/1126-6708/2007/04/008}{\emph{JHEP} {\bf 04}
  (2007) 008}, [\href{http://arxiv.org/abs/hep-th/0612168}{{\tt
  hep-th/0612168}}].

\bibitem{Buchbinder:2014nia}
E.~I. Buchbinder and A.~A. Tseytlin, \emph{{1/N correction in the D3-brane
  description of a circular Wilson loop at strong coupling}},
  \href{http://dx.doi.org/10.1103/PhysRevD.89.126008}{\emph{Phys. Rev.} {\bf
  D89} (2014) 126008}, [\href{http://arxiv.org/abs/1404.4952}{{\tt
  1404.4952}}].

\bibitem{Chen-Lin:2016kkk}
X.~Chen-Lin, \emph{{Symmetric Wilson Loops beyond leading order}},
  \href{http://dx.doi.org/10.21468/SciPostPhys.1.2.013}{\emph{SciPost Phys.}
  {\bf 1} (2016) 013}, [\href{http://arxiv.org/abs/1610.02914}{{\tt
  1610.02914}}].

\bibitem{Marino:2004eq}
M.~Marino, \emph{{Les Houches lectures on matrix models and topological
  strings}},  in \emph{{Les Houches School on Applications of Random Matrices
  in Physics}}, 2004.
\newblock \href{http://arxiv.org/abs/hep-th/0410165}{{\tt hep-th/0410165}}.

\bibitem{Wadia:2012fr}
S.~R. Wadia, \emph{{A Study of U(N) Lattice Gauge Theory in 2-dimensions {\rm
  (edited version of an unpublished 1979 EFI (U. Chicago) preprint)}}},
  \href{http://arxiv.org/abs/1212.2906}{{\tt 1212.2906}}.

\bibitem{Brezin:1992yc}
E.~Brezin and J.~Zinn-Justin, \emph{{Renormalization group approach to matrix
  models}}, \href{http://dx.doi.org/10.1016/0370-2693(92)91953-7}{\emph{Phys.
  Lett.} {\bf B288} (1992) 54--58},
  [\href{http://arxiv.org/abs/hep-th/9206035}{{\tt hep-th/9206035}}].

\bibitem{Higuchi:1994dv}
S.~Higuchi, C.~Itoi, S.~Nishigaki and N.~Sakai, \emph{{Large N renormalization
  group approach to matrix models}},  in \emph{{Group theoretical methods in
  physics. Proceedings, 40th Yamada Conference, 20th International Colloquium,
  Toyonaka, Japan, July 4-9, 1994}}, 1994.
\newblock \href{http://arxiv.org/abs/hep-th/9409157}{{\tt hep-th/9409157}}.

\bibitem{Periwal:1990gf}
V.~Periwal and D.~Shevitz, \emph{{Unitary Matrix Models as Exactly Solvable
  String Theories}},
  \href{http://dx.doi.org/10.1103/PhysRevLett.64.1326}{\emph{Phys. Rev. Lett.}
  {\bf 64} (1990) 1326}.

\bibitem{Periwal:1990qb}
V.~Periwal and D.~Shevitz, \emph{{Exactly Solvable Unitary Matrix Models:
  Multicritical Potentials and Correlations}},
  \href{http://dx.doi.org/10.1016/0550-3213(90)90676-5}{\emph{Nucl. Phys.} {\bf
  B344} (1990) 731--746}.

\bibitem{Klebanov:2003wg}
I.~R. Klebanov, J.~M. Maldacena and N.~Seiberg, \emph{{Unitary and complex
  matrix models as 1-d type 0 strings}},
  \href{http://dx.doi.org/10.1007/s00220-004-1183-7}{\emph{Commun. Math. Phys.}
  {\bf 252} (2004) 275--323}, [\href{http://arxiv.org/abs/hep-th/0309168}{{\tt
  hep-th/0309168}}].

\bibitem{Liu:2004vy}
H.~Liu, \emph{{Fine structure of Hagedorn transitions}},
  \href{http://arxiv.org/abs/hep-th/0408001}{{\tt hep-th/0408001}}.

\bibitem{AlvarezGaume:2005fv}
L.~Alvarez-Gaume, C.~Gomez, H.~Liu and S.~Wadia, \emph{{Finite temperature
  effective action, AdS(5) black holes, and 1/N expansion}},
  \href{http://dx.doi.org/10.1103/PhysRevD.71.124023}{\emph{Phys. Rev.} {\bf
  D71} (2005) 124023}, [\href{http://arxiv.org/abs/hep-th/0502227}{{\tt
  hep-th/0502227}}].

\bibitem{Witten:1998zw}
E.~Witten, \emph{{Anti-de Sitter space, thermal phase transition, and
  confinement in gauge theories}}, {\emph{Adv. Theor. Math. Phys.} {\bf 2}
  (1998) 505--532}, [\href{http://arxiv.org/abs/hep-th/9803131}{{\tt
  hep-th/9803131}}].

\bibitem{Marino:2008ya}
M.~Marino, \emph{{Nonperturbative effects and nonperturbative definitions in
  matrix models and topological strings}},
  \href{http://dx.doi.org/10.1088/1126-6708/2008/12/114}{\emph{JHEP} {\bf 12}
  (2008) 114}, [\href{http://arxiv.org/abs/0805.3033}{{\tt 0805.3033}}].

\bibitem{shenker1991strength}
S.~H. Shenker, \emph{The strength of nonperturbative effects in string theory},
   in \emph{Random surfaces and quantum gravity}, pp.~191--200.
\newblock Springer, 1991.

\bibitem{David:1990sk}
F.~David, \emph{{Phases of the large N matrix model and nonperturbative effects
  in 2-d gravity}},
  \href{http://dx.doi.org/10.1016/0550-3213(91)90202-9}{\emph{Nucl. Phys.} {\bf
  B348} (1991) 507--524}.

\bibitem{Buividovich:2015oju}
P.~V. Buividovich, G.~V. Dunne and S.~N. Valgushev, \emph{{Complex Path
  Integrals and Saddles in Two-Dimensional Gauge Theory}},
  \href{http://dx.doi.org/10.1103/PhysRevLett.116.132001}{\emph{Phys. Rev.
  Lett.} {\bf 116} (2016) 132001}, [\href{http://arxiv.org/abs/1512.09021}{{\tt
  1512.09021}}].

\bibitem{Alvarez:2016rmo}
G.~Alvarez, L.~Martinez~Alonso and E.~Medina, \emph{{Complex saddles in the
  Gross-Witten-Wadia matrix model}},
  \href{http://dx.doi.org/10.1103/PhysRevD.94.105010}{\emph{Phys. Rev.} {\bf
  D94} (2016) 105010}, [\href{http://arxiv.org/abs/1610.09948}{{\tt
  1610.09948}}].

\bibitem{Okuyama:2017pil}
K.~Okuyama, \emph{{Wilson loops in unitary matrix models at finite $N$}},
  \href{http://dx.doi.org/10.1007/JHEP07(2017)030}{\emph{JHEP} {\bf 07} (2017)
  030}, [\href{http://arxiv.org/abs/1705.06542}{{\tt 1705.06542}}].

\bibitem{Ambjorn:1992gw}
J.~Ambjorn, L.~Chekhov, C.~F. Kristjansen and {\relax Yu}.~Makeenko,
  \emph{{Matrix model calculations beyond the spherical limit}},
  \href{http://dx.doi.org/10.1016/0550-3213(93)90476-6,
  10.1016/0550-3213(95)00391-5}{\emph{Nucl. Phys.} {\bf B404} (1993) 127--172},
  [\href{http://arxiv.org/abs/hep-th/9302014}{{\tt hep-th/9302014}}].

\bibitem{Mizoguchi:2004ne}
S.~Mizoguchi, \emph{{On unitary / hermitian duality in matrix models}},
  \href{http://dx.doi.org/10.1016/j.nuclphysb.2005.03.035}{\emph{Nucl. Phys.}
  {\bf B716} (2005) 462--486}, [\href{http://arxiv.org/abs/hep-th/0411049}{{\tt
  hep-th/0411049}}].

\bibitem{Bessis:1979is}
D.~Bessis, \emph{{A New Method in the Combinatorics of the Topological
  Expansion}}, \href{http://dx.doi.org/10.1007/BF01221445}{\emph{Commun. Math.
  Phys.} {\bf 69} (1979) 147}.

\bibitem{Bessis:1980ss}
D.~Bessis, C.~Itzykson and J.~B. Zuber, \emph{{Quantum field theory techniques
  in graphical enumeration}},
  \href{http://dx.doi.org/10.1016/0196-8858(80)90008-1}{\emph{Adv. Appl. Math.}
  {\bf 1} (1980) 109--157}.

\bibitem{DiFrancesco:1993cyw}
P.~Di~Francesco, P.~H. Ginsparg and J.~Zinn-Justin, \emph{{2-D Gravity and
  random matrices}},
  \href{http://dx.doi.org/10.1016/0370-1573(94)00084-G}{\emph{Phys. Rept.} {\bf
  254} (1995) 1--133}, [\href{http://arxiv.org/abs/hep-th/9306153}{{\tt
  hep-th/9306153}}].

\bibitem{Goldschmidt:1979hq}
Y.~Y. Goldschmidt, \emph{{1/$N$ Expansion in Two-dimensional Lattice Gauge
  Theory}}, \href{http://dx.doi.org/10.1063/1.524600}{\emph{J. Math. Phys.}
  {\bf 21} (1980) 1842}.

\bibitem{Bars:1979xb}
I.~Bars and F.~Green, \emph{{Complete Integration of U ($N$) Lattice Gauge
  Theory in a Large $N$ Limit}},
  \href{http://dx.doi.org/10.1103/PhysRevD.20.3311}{\emph{Phys. Rev.} {\bf D20}
  (1979) 3311}.

\bibitem{Makeenko:1979pb}
{\relax Yu}.~M. Makeenko and A.~A. Migdal, \emph{{Exact Equation for the Loop
  Average in Multicolor QCD}},
  \href{http://dx.doi.org/10.1016/0370-2693(79)90131-X}{\emph{Phys. Lett.} {\bf
  88B} (1979) 135}.

\bibitem{Migdal:1984gj}
A.~A. Migdal, \emph{{Loop Equations and 1/N Expansion}},
  \href{http://dx.doi.org/10.1016/0370-1573(83)90076-5}{\emph{Phys. Rept.} {\bf
  102} (1983) 199--290}.

\bibitem{Paffuti:1980cs}
G.~Paffuti and P.~Rossi, \emph{{A Solution of Wilson's Loop Equation in Lattice
  {QCD} in Two-dimensions}},
  \href{http://dx.doi.org/10.1016/0370-2693(80)90273-7}{\emph{Phys. Lett.} {\bf
  92B} (1980) 321--323}.

\bibitem{Wadia:1980rb}
S.~R. Wadia, \emph{{On the Dyson-schwinger Equations Approach to the Large $N$
  Limit: Model Systems and String Representation of {Yang-Mills} Theory}},
  \href{http://dx.doi.org/10.1103/PhysRevD.24.970}{\emph{Phys. Rev.} {\bf D24}
  (1981) 970}.

\bibitem{Friedan:1980tu}
D.~Friedan, \emph{{Some Nonabelian Toy Models in the Large $N$ Limit}},
  \href{http://dx.doi.org/10.1007/BF01942328}{\emph{Commun. Math. Phys.} {\bf
  78} (1981) 353}.

\bibitem{Green:1980bg}
F.~Green and S.~Samuel, \emph{{Chiral Models: Their Implication for Gauge
  Theories and Large $N$}},
  \href{http://dx.doi.org/10.1016/0550-3213(81)90486-7}{\emph{Nucl. Phys.} {\bf
  B190} (1981) 113--150}.

\bibitem{Akemann:2001st}
G.~Akemann and P.~H. Damgaard, \emph{{Wilson loops in $N$=4 supersymmetric
  Yang-Mills theory from random matrix theory}},
  \href{http://dx.doi.org/10.1016/S0370-2693(01)00675-X,
  10.1016/S0370-2693(01)01346-6}{\emph{Phys. Lett.} {\bf B513} (2001) 179},
  [\href{http://arxiv.org/abs/hep-th/0101225}{{\tt hep-th/0101225}}].

\bibitem{Marino:2007te}
M.~Marino, R.~Schiappa and M.~Weiss, \emph{{Nonperturbative Effects and the
  Large-Order Behavior of Matrix Models and Topological Strings}},
  \href{http://dx.doi.org/10.4310/CNTP.2008.v2.n2.a3}{\emph{Commun. Num. Theor.
  Phys.} {\bf 2} (2008) 349--419}, [\href{http://arxiv.org/abs/0711.1954}{{\tt
  0711.1954}}].

\bibitem{Makeenko:1991tb}
{\relax Yu}.~Makeenko, \emph{{Loop equations in matrix models and in 2-D
  quantum gravity}},
  \href{http://dx.doi.org/10.1142/S0217732391002050}{\emph{Mod. Phys. Lett.}
  {\bf A6} (1991) 1901--1913}.

\bibitem{Gross:1989vs}
D.~J. Gross and A.~A. Migdal, \emph{{Nonperturbative Two-Dimensional Quantum
  Gravity}}, \href{http://dx.doi.org/10.1103/PhysRevLett.64.127}{\emph{Phys.
  Rev. Lett.} {\bf 64} (1990) 127}.

\bibitem{Gross:1989aw}
D.~J. Gross and A.~A. Migdal, \emph{{A Nonperturbative Treatment of
  Two-dimensional Quantum Gravity}},
  \href{http://dx.doi.org/10.1016/0550-3213(90)90450-R}{\emph{Nucl. Phys.} {\bf
  B340} (1990) 333--365}.

\bibitem{Karczmarek:2010ec}
J.~L. Karczmarek, G.~W. Semenoff and S.~Yang, \emph{{Comments on k-Strings at
  Large N}}, \href{http://dx.doi.org/10.1007/JHEP03(2011)075}{\emph{JHEP} {\bf
  03} (2011) 075}, [\href{http://arxiv.org/abs/1012.5875}{{\tt 1012.5875}}].

\bibitem{Grignani:2009ua}
G.~Grignani, J.~L. Karczmarek and G.~W. Semenoff, \emph{{Hot Giant Loop
  Holography}}, \href{http://dx.doi.org/10.1103/PhysRevD.82.027901}{\emph{Phys.
  Rev.} {\bf D82} (2010) 027901}, [\href{http://arxiv.org/abs/0904.3750}{{\tt
  0904.3750}}].

\end{thebibliography}\endgroup
\bibliographystyle{JHEP}

\end{document}